\documentclass[twocolumn,showpacs,preprintnumbers,amsmath,amssymb]{revtex4}

\usepackage{amsmath,amssymb}

\usepackage{color}
\usepackage[dvips]{graphicx}
\usepackage{subfigure}

\newcommand{\lsim}{\mathrel{\mathop{\kern 0pt \rlap
  {\raise.2ex\hbox{$<$}}}
  \lower.9ex\hbox{\kern-.190em $\sim$}}}
\newcommand{\gsim}{\mathrel{\mathop{\kern 0pt \rlap
  {\raise.2ex\hbox{$>$}}}
  \lower.9ex\hbox{\kern-.190em $\sim$}}}
\newcommand{\sigmav}{\langle \sigma_{\rm ann} v \rangle}

\newcommand{\beq}{\begin{equation}}
\newcommand{\eeq}{\end{equation}}
\newcommand{\bea}{\begin{eqnarray}}
\newcommand{\ena}{\end{eqnarray}}
\newcommand{\etal}{{\it et al.}}

\renewcommand{\prd}[3]{Phys.\ Rev.\ D\ {\bf #1}, #3 (#2)}

\begin{document}

\preprint{DFTT 2/2004}

\title{Indirect signals from light neutralinos\\ 
in supersymmetric models without gaugino mass unification}



%
\author{A. Bottino}
\affiliation{Dipartimento di Fisica Teorica, Universit\`a di Torino \\
Istituto Nazionale di Fisica Nucleare, Sezione di Torino \\
via P. Giuria 1, I--10125 Torino, Italy}

\author{F. Donato}
\affiliation{Dipartimento di Fisica Teorica, Universit\`a di Torino \\
Istituto Nazionale di Fisica Nucleare, Sezione di Torino \\
via P. Giuria 1, I--10125 Torino, Italy}

\author{N. Fornengo}
\affiliation{Dipartimento di Fisica Teorica, Universit\`a di Torino \\
Istituto Nazionale di Fisica Nucleare, Sezione di Torino \\
via P. Giuria 1, I--10125 Torino, Italy}

\author{S. Scopel}
\email{bottino@to.infn.it, donato@to.infn.it, fornengo@to.infn.it, scopel@to.infn.it}
\homepage{http://www.astroparticle.to.infn.it}
\affiliation{Dipartimento di Fisica Teorica, Universit\`a di Torino \\
Istituto Nazionale di Fisica Nucleare, Sezione di Torino \\
via P. Giuria 1, I--10125 Torino, Italy}

\date{\today}

\begin{abstract}
We examine indirect signals produced by neutralino
self--annihilations, in the galactic halo or inside celestial
bodies, in the frame of an effective MSSM model without gaugino-mass
unification at a grand unification scale. 
We compare our theoretical predictions with current
experimental data of gamma--rays and antiprotons in space and of
upgoing muons at neutrino telescopes. 
Results are presented for a wide range of the neutralino mass, 
though our discussions are focused on light neutralinos.
We find that only the antiproton
signal is potentially able to set constraints on very low--mass
neutralinos, below 20 GeV. The gamma--ray signal, both from the
galactic center and from high galactic latitudes, requires
significantly steep profiles or substantial clumpiness in order to
reach detectable levels. The up-going muon signal is largely below
experimental sensitivities for the neutrino flux coming from the Sun;
for the flux from the Earth an improvement of about one order of
magnitude in experimental sensitivities (with a low energy threshold)
can make accessible neutralino masses close to $O$, $Si$ and
$Mg$ nuclei masses, for which resonant capture is operative.
\end{abstract}


\maketitle

\section{Introduction}
\label{sec:intro}
 
In Supersymmetric models without gaugino-mass unification at the
grand unification scale, neutralinos can be lighter than the current
lower bound of 50 GeV, which instead occurs in the case of
gaugino--universal models. In Refs. \cite{lowneu,lowmass} we discussed
the properties of these light neutralinos as relic particles
($R$--parity conservation is assumed) and showed that an absolute
lower limit of 6 GeV on the neutralino mass $m_{\chi}$ can be placed
by applying the most recent determinations of the upper bound on the
Cold Dark Matter (CDM) content in the Universe, in combination with
constraints imposed on the Higgs and supersymmetric parameters by
measurements at colliders and other precision experiments, like the
muon anomalous magnetic moment and the rare decay $b \rightarrow s +
\gamma$. In Refs. \cite{lowneu,lowmass} we also showed that direct
detection rates for light relic neutralinos make these particles
detectable with WIMP direct search experiments with current
technologies. A comparison of our predictions with intervening
experimental results was presented in Ref. \cite{lowdir} .

In the present paper we examine light neutralinos in connection with
the indirect signals which can be produced by neutralino
self--annihilations in the galactic halo or inside celestial
bodies. We compare our theoretical predictions with current
experimental data on measurements of gamma--rays and antiprotons in
space and of up-going muons at neutrino telescopes.  Results are
presented for a wide range of the neutralino mass, from the
established lower bound of 6 GeV up to 500 GeV. However, our
discussions are focused on light neutralinos (i.e. neutralinos with
$m_{\chi} \lsim$ 50 GeV), since these are not usually considered in
current literature \cite{otherlow}. 

The structure of the paper is the following. In Sec. \ref{sec:susy} we
briefly summarize the gaugino non--universal supersymmetric model and
the properties of light neutralinos which arise in this framework. In
Sec. \ref{sec:dmdf} we discuss the dark matter density distribution in
the galactic halo, which is relevant to indirect detection signals,
especially to the gamma--ray flux. In Sec. \ref{sec:gamma} we present
the calculation and comparison with data of the gamma--ray flux: we
consider both the signal coming from the galactic center and from high
galactic latitudes. In Sec. \ref{sec:pbar} we discuss the antiproton
signal, whereas in Sec. \ref{sec:nu} we show our results for the
indirect signals at neutrino telescopes.
Our conclusions are drawn in Sec. \ref{sec:conclusions}.

\section{Supersymmetric models without gaugino--mass unification}
\label{sec:susy}

A typical assumption of supersymmetric models is the unification
condition for the three gaugino masses $M_{1,2,3}$ at the GUT scale:
$M_1=M_2=M_3$. This hypothesis implies that at the electroweak scale
$M_1 \simeq 0.5~M_2$. Under this unification condition the bound on
the neutralino mass is determined to be $m_{\chi} \gsim 50$ GeV. This
is derived from the lower bound on the chargino mass (which depends on
$M_2$ but not on $M_1$) determined at LEP2: $m_{\chi^\pm} \gsim$ 100
GeV. By allowing a deviation from gaugino--universality, the
neutralino can be lighter than in the gaugino--universal models when
$M_1 \equiv R~M_2$, with $R<0.5$. In this case current data from
accelerators do not set an absolute lower bound on $m_\chi$.

We consider here an extension of the MSSM which allows for a deviation
from gaugino--mass universality by the introduction of the parameter
$R$, varied here in the interval: $(0.01 \div 0.5)$. This range for
$R$ implies that the accelerator lower bound on the neutralino mass
can be moved down to few GeV for $R\sim 0.01$. The ensuing light
neutralinos have a dominant bino component; a deviation from a pure
bino composition is mainly due to a mixture of $\tilde B$ with $\tilde
H_1^{\circ}$ \cite{lowneu,lowmass,lowdir}. Notice that our range of
$R$ includes also the usual model with gaugino--mass universality.

We therefore employ an effective MSSM scheme at the electroweak scale,
defined in terms of a minimal number of parameters, only those
necessary to shape the essentials of the theoretical structure of MSSM
and of its particle content, supplemented by the gaugino
non--universality parameter $R$. The assumptions that we impose at the
electroweak scale are: a) all squark soft--mass parameters are
degenerate: $m_{\tilde q_i} \equiv m_{\tilde q}$; b) all slepton
soft--mass parameters are degenerate: $m_{\tilde l_i} \equiv m_{\tilde
l}$; c) all trilinear parameters vanish except those of the third
family, which are defined in terms of a common dimensionless parameter
$A$: $A_{\tilde b} = A_{\tilde t} \equiv A m_{\tilde q}$ and
$A_{\tilde \tau} \equiv A m_{\tilde l}$.  As a consequence, the
supersymmetric parameter space consists of the following independent
parameters: $M_2, \mu, \tan\beta, m_A, m_{\tilde q}, m_{\tilde l}, A$
and $R$. In the previous list of parameters we have denoted by $\mu$
the Higgs mixing mass parameter, by $\tan\beta$ the ratio of the two
Higgs v.e.v.'s and by $m_A$ the mass of the CP-odd neutral Higgs
boson.

In the numerical random scanning of the supersymmetric parameter space
we have used the following ranges: $1 \leq \tan \beta \leq 50$, $100~
{\rm GeV }\leq |\mu|$, $M_2 \leq 1000~{\rm GeV }$, $100~{\rm GeV}\leq
m_{\tilde q}, m_{\tilde l} \leq 1000~{\rm GeV }$, ${\rm
sign}(\mu)=-1,1$, $90~{\rm GeV }\leq m_A \leq 1000~{\rm GeV }$, $-3
\leq A \leq 3$, in addition to the above mentioned range $0.01 \leq R
\leq 0.5$. We impose the experimental constraints: accelerators data
on supersymmetric and Higgs boson searches and on the invisible width of the Z
boson, measurements of the branching ratio of the
$b\rightarrow s + \gamma$ decay and of the upper bound on the branching ratio
of $B_s \rightarrow \mu^+ + \mu^-$, and measurements of the muon anomalous magnetic
moment $a_\mu \equiv (g_{\mu} - 2)/2$. The range used here for the $b
\rightarrow s + \gamma$ branching ratio is $2.18 \cdot 10^{-4} \leq
BR (b \rightarrow s + \gamma) \leq 4.28 \cdot 10^{-4}$
 \cite{bsgamma}. For the branching ratio of $B_s \rightarrow \mu^+ + \mu^-$
 we employ the upper limit $BR(B_s \rightarrow \mu^+ + \mu^-) < 7.5 \cdot 10^{-7}$ 
(95\% C.L.) \cite{bsmumu_exp}; for the theoretical evaluation we have used the
results of \cite{bsmumu_th} with inclusion of the QCD radiative corrections 
to the $b$ Yukawa coupling \cite{carena}. 
 For the deviation of the current experimental world average of $a_{\mu}$ from
the theoretical evaluation within the Standard Model we use the 2$\sigma$
range: $-142 \leq \Delta a_{\mu} \cdot 10^{11} \leq 474 $; this interval takes
into account the recent evaluations of Refs. \cite{davier,hagiwara2}. We notice
that gluinos do not enter directly into our loop contributions to $BR(b \rightarrow s +
\gamma)$ and $BR(B_s \rightarrow \mu^+ + \mu^-)$, since we assume flavor-diagonal
sfermion mass matrices. Thus, gluinos appear only in the QCD radiative
corrections to the $b$ Yukawa coupling;  in the evaluation of these effects the
value of the relevant mass parameter $M_3$ is taken at the standard unification
value $M_3 = M_2 \; \alpha_3(M_Z)/\alpha_2(M_Z)$, where
$\alpha_3(M_Z)$ and $\alpha_2(M_Z)$ are the SU(3) and SU(2) coupling
constants evaluated at the scale $M_Z$.

The new data on the cosmic microwave background from WMAP\cite{wmap},
used in combination with other cosmological observations, mainly
galaxy surveys and Lyman--$\alpha$ forest data, are sharpening our
knowledge of the cosmological parameters, and in particular of the
amount of dark matter in the Universe.  From the analysis of
Ref. \cite{wmap}, we obtain a restricted range for the relic density
of a cold species like the neutralinos. The density parameter of cold
dark matter is bounded at $2\sigma$ level by the values:
$(\Omega_{CDM} h^2)_{\rm min} = 0.095$ and $(\Omega_{CDM} h^2)_{\rm
max} = 0.131$. This is the range for CDM that we consider in the
present paper. An independent determination for the content of cold
dark matter in the Universe is provided by the Sloan Digital Sky
Survey Collaboration \cite{sloan}; this new data agrees with the
results of Ref. \cite{wmap}.

We recall that the relic abundance $\Omega_{\chi} h^2$ is essentially
given by $\Omega_{\chi} h^2 \propto <\sigma_{\rm ann} v>^{-1}_{\rm
int}$, where $<\sigma_{\rm ann} v>_{\rm int}$ is the
thermally--averaged neutralino annihilation cross section times
average velocity, integrated from the freeze--out temperature $T_f$ to
the present one $T_0$. The quantity $\sigma_{\rm ann}$ enters also in
the calculation of the indirect signals that will be discussed in the
following sections. In the evaluation of $\sigma_{\rm ann}$ we have
considered the full set of available final states: fermion-antifermion
pairs, gluon pairs, pairs of charged Higgs bosons, one Higgs boson and
one gauge boson, pairs of gauge bosons \cite{anni}. 
We have not included coannihilation in our evaluation of the  neutralino
relic abundance, since, at variance from a constrained SUGRA scheme, in our
effective supersymmetric model a matching of the neutralino mass with other
masses is accidental, i.e. not induced by some intrinsic relationship among
different parameters.
Introducing coannihilation would only produce an insignificant reshuffle in the
representative points of the scatter plots displayed in the present paper,
without a modification of their borders, which are the only feature of physical
significance.

The relic abundance $\Omega_\chi h^2$ of neutralinos lighter than 50
GeV which arise in our class of gaugino non--universal models has a
relatively simple structure in terms of dominant diagrams in the
annihilation cross section \cite{lowneu,lowmass}. Here we just
remind that combining our calculation of the relic
abundance of light neutralinos with the value of $(\Omega_{CDM}
h^2)_{\rm max}$, an absolute lower bound on the neutralino mass of 6
GeV can be set \cite{lowneu,lowmass}. 
We note that within our present scanning of the supersymmetric parameter
space, the lower limit on the neutralino mass shifts to about 7 GeV,  when the
upper bound on $BR(B_s \rightarrow \mu^+ + \mu^-) < 7.5 \cdot 10^{-7}$ is implemented 
(this constraint was not included in \cite{lowneu,lowmass}).
It is remarkable that a lower limit on $m_\chi$ is set
not by searches at accelerators, but instead by cosmological
arguments.

\section{Dark matter in the Galaxy}
\label{sec:dmdf}

Signals due to neutralino self--annihilation in the halo depend
quadratically on the dark matter density distribution $\rho(\vec r)$,
and are therefore very sensitive to the features of this physical
quantity. Two properties are of special relevance: 1) the behavior
of the density distribution in the galactic center (GC), 2) the extent
of the density contrast (clumpiness), which represents the deviation
of the actual density distribution from a smooth distribution.

The most commonly used density distributions can be parametrized by
the following spherically--averaged density profile \cite{hernquist}:
\begin{equation}
\rho(r) = \rho_l \left(\frac{r_\odot}{r}\right)^\gamma
\left[\frac{1+(r_\odot/a)^\alpha}{1+(r/a)^\alpha}\right]^{(\beta
-\gamma)/\alpha},  
\label{eq:density}
\end{equation}
where $r=|\vec r|$, $r_\odot=8$ kpc \cite{S2} is the distance of the
Sun from the galactic center, $a$ is a scale length and $\rho_l$ is
the total local (solar neighborhood) dark matter density. In
particular, the isothermal density profile corresponds to $(\alpha,
\beta, \gamma)$ = (2,2,0), the Navarro, Frenk and White (NFW) profile
\cite{nfw} corresponds to $(\alpha, \beta,\gamma)$ = (1,3,1) and the
Moore et al.  profile \cite{moore} to $(\alpha,\beta, \gamma)$ =
(1.5,3,1.5). The two latter profiles, both derived from numerical
simulations of structure formation, differ noticeably in their
behavior at small distances from the GC: $r^{-1}$ for the NFW and
$r^{-1.5}$ for the Moore et al. profile, with ensuing large
differences in the size of the expected signals for WIMP annihilation
from the central region of the Galaxy.

Recent results of extensive numerical simulations, aimed at an
analysis of the inner structure of $\Lambda CDM$ halos, strongly
disfavor a behavior as singular as $r^{-1.5}$, but also indicate
that a NFW profile is likely not to be adequate at small distances
from the GC \cite{navarro}. It turns out that the density profile is
not described by a singular power--law at small distances; rather, in
this asymptotic regime, the numerical results are well fitted by a
profile whose logarithmic slope $\delta(r) \equiv - d(\ln \;
\rho(r))/d(\ln \; r)$ is given by $\delta(r) \sim r^{-\alpha}$ with
$\alpha \simeq 0.17$. This leads to a non--singular dark matter
density distribution function of the form \cite{navarro}:
\begin{equation}
 \rho(r) = \rho_{-2} \; \exp\left\{ -\frac{2}{\alpha}
 \left[\left(\frac{r}{r_{-2}}\right)^\alpha -1\right] \right\},
 \label{eq:alpha}
\end{equation}
where $r_{-2}$ is the radius where the logarithmic slope is $\delta =
-2$, and $\rho_{-2} \equiv \rho(r_{-2})$.  These various distributions
mainly differ in their behavior at small $r$.  We wish to stress here
that anyway current cosmological simulations are not reliable for
radii smaller than an $r_{\rm min}\simeq 0.1$ -- 1 kpc. We also notice
that singular profiles are subject of debate in current literature,
with analyses pointing to inconsistencies with observational data on
rotational curves \cite{salucci}.

Furthermore we recall that other density profiles are able to describe
the dark matter halo, including for instance different classes of
logarithmic and power--law potentials, axisymmetric distributions or
even triaxial distributions \cite{bcfs}. In the following we will
concentrate on the standard isotropic density profiles of
Eq. (\ref{eq:density}) and on the new profile of Eq. (\ref{eq:alpha})
deduced from numerical simulations. For definiteness, we use as a
reference model the NFW density distribution, and discuss the
deviation from this reference case when the other density profiles are
considered. Most signals do not depend or are only mildly dependent on
the critical behavior of the dark matter density around the GC: this
occurs for the neutrino fluxes from the Earth and the Sun, for
antiprotons and for gamma--rays coming from large galactic
latitudes. On the contrary, gamma-rays coming from the GC are very
sensitive to the inner parts of the Galaxy and the differences between
the different halo profiles will be explicitly discussed.

A key parameter for all the density distributions is the value for the
total local dark matter density $\rho_l$. This parameter can be
determined for each density profile assuming compatibility with the
measurements of rotational curves and the total mass of the Galaxy
\cite{bcfs}. For instance, a simple modelling of the visible and dark
components of the Galaxy showed that $\rho_l$ can range from 0.18 GeV
cm$^{-3}$ to 0.71 GeV cm$^{-3}$ for an isothermal sphere profile, from
0.20 GeV cm$^{-3}$ to 1.11 GeV cm$^{-3}$ for a NFW distribution and
0.22 GeV cm$^{-3}$ to 0.98 GeV cm$^{-3}$ for a Moore et al. shape
\cite{bcfs}. For definiteness, our results will be presented for
$\rho_l = 0.3$ GeV cm$^{-3}$ for all the density profiles employed in
the present analysis. The parameter $\rho_l$ enters as a mere scaling
factor in the signal fluxes: the effect of varying $\rho_l$ is
therefore easily taken into account.

Once the density profile which describes the total dark matter density
in the galactic halo is chosen, the actual neutralino density
distribution is taken to be:
\begin{equation}
\rho_{\chi}(r) = \xi \rho(r),  
\label{eq:rescal}
\end{equation}
where $\xi$ accounts for the fact that neutralinos could be only a
fraction of the total cold dark matter ($\xi \leq 1$). This
characteristic is linked to the actual relic abundance of neutralinos
and is accounted for by using the standard rescaling prescription:
$\xi = \min[1,\Omega_{\chi} h^2/(\Omega_{CDM} h^2)_{\rm min}]$.

As was noticed in Refs. \cite{bss,ss,bbm} an effect of density
contrast in the dark matter distribution could produce a strong
enhancement effect in signals due to $\chi$-$\chi$ annihilations in
the halo. This property was subsequently considered in connection with
various signals (gamma rays, positrons, antiprotons)
\cite{clump,clump2}, sometimes under the assumption of a strong
clumpiness effect, at the level of a few orders of magnitude. However,
according to a recent analytical investigation on the production of
small-scale dark matter clumps \cite{bde}, the clumpiness effect would
not be large, with the result that the ensuing enhancement on the
annihilation signals is limited to a factor of a few. Similar
conclusions are also reached in Ref.  \cite{setal} by using results of
high-resolution numerical simulations.
 
\section{Gamma rays}
\label{sec:gamma}

The flux of gamma rays $\Phi_\gamma(E_\gamma, \psi)$ originated from
neutralino pair annihilation in the galactic halo
\cite{bss,clump,gamma_history} and coming from the angular direction
$\psi$ is given by:
\begin{equation}
\Phi_\gamma(E_\gamma, \psi) = \frac{1}{4\pi} \frac{\langle\sigma_{\rm
ann} v\rangle}{m_\chi^2} \frac{dN_\gamma}{d E_{\gamma}}
\frac{1}{2}I(\psi) \, ,
\label{eq:flux_gamma}
\end{equation}
where $\sigmav$ is the annihilaton cross section times the relative
velocity mediated over the galactic velocity distribution function and
$dN_\gamma/d E_{\gamma}$ is the energy spectrum of $\gamma$-rays
originated from a single neutralino pair annihilation. The quantity
$I(\psi)$ is the integral of the squared dark matter density
distribution performed along the line of sight:
\begin{equation}
I(\psi) = \int_{\rm l.o.s} \rho^2(r(\lambda,\psi))~ d\lambda(\psi),
\label{eq:los}
\end{equation}
and $\psi$ is the angle between the line of sight and the line
pointing toward the GC. The angle $\psi$ is related to the galactic
longitude $l$ and latitude $b$ by the expression $\cos\psi = \cos l
\cos b$. A point at a distance $\lambda$ from us and observed under an
angle $\psi$ is therefore located at the galactocentric distance $r =
\sqrt{\lambda^2 + r_\odot^2 - 2 \lambda r_\odot \cos\psi}$. The factor
of $1/2$ in Eq.(\ref{eq:flux_gamma}) is due to the fact that the
gamma--ray flux depends on the number of neutralino pairs present in
the galactic halo, as pointed out in Ref. \cite{clump1}. This factor
of $1/2$ applies as well to any other indirect detection signal which
depends on the annihilation of a pair of Majorana fermions in the
galactic halo, like positrons, antiprotons, antideuterium. In the case
of dark matter composed of Dirac fermions, the statistical factor
would instead be $1/4$, if $\rho(r)$ describes the total dark
matter distribution ascribed to the given Dirac species.

\subsection{The source spectrum}
\label{sec:source_gamma}

As far as the annihilation of light neutralinos is concerned (namely
for neutralino masses below the thresholds for gauge--bosons,
higgs-bosons and $t$ quark production), the production of
$\gamma$--rays in the continuum takes contribution mainly from the
hadronization of quarks and gluon pairs produced in the neutralino
annihilation process. The subsequent $\pi^0$ production and decay
$\pi^0 \longrightarrow 2\gamma$ give usually the dominant
contribution. In this case the $\gamma$--ray energy spectrum is given
by:
\begin{equation}
\frac{d N_{\gamma}}{d E_{\gamma}}=\int_{E_{\pi}^{\rm min}}^{E_{\pi}^{\rm max}}
P(E_{\pi},E_{\gamma}) \frac{d N_{\pi}}{d E_{\pi}} d E_{\pi},
\label{eq:pi0}
\end{equation}
where $P(E_{\pi},E_{\gamma})=2(E_{\pi}^2-m_{\pi}^2)^{-1/2}$ is the
probability per unit energy to produce a $\gamma$--ray with energy
$E_{\gamma}$ out of a pion with energy $E_{\pi}$, while $d N_{\pi}/d
E_{\pi}$ is the pion yield per annihilation event.

We have evaluated the quantity $d N_{\pi}/d E_{\pi}$ by means of a
Monte Carlo simulation with the PYTHIA package \cite{pythia}. The
Monte Carlo code has been run by injecting $q\bar{q}$ and gluon pairs
back--to--back at fixed center--of--mass energy $E_{\rm cm}=2 \,
m_{\chi}$. Since quarks and gluons are confined, they contribute to a
complex final--state pattern of out--going hadronic strings decaying
to physical hadrons through fragmentation. In the Lund string scheme,
fragmentation is an intrinsically scale invariant process. This
implies that, in the rest frame of the decaying string, the
final--state spectrum is invariant in the variable $x\equiv E_F/m_{\rm
string}$, where $E_F$ is the energy of the given final state and
$m_{\rm string}$ is the total string mass. Were it not for showered
gluons, a $q\bar{q}$ pair from neutralino annihilation would produce,
in the reference frame of the two annihilating neutralinos, a single
hadronic string at rest with $m_{\rm string}=2 \, m_{\chi}$,
subsequently fragmenting to produce (among other particles) a pion
spectrum which would be scale--invariant in the variable $y \equiv
E_{\pi}/m_{\chi}$. However, due to showering, this scale invariance is
significantly broken, since the pion energy--spectrum is given by the
superposition of different decaying strings boosted at different
energies. Therefore the pion spectrum at a given neutralino mass
cannot be obtained from that calculated at a different $m_{\chi}$.

We have therefore evaluated the pion yield per annihilation event,
$dN_{\pi}^F(m_{\chi},E_{\pi})/d E_{\pi}$, for each final state
$F=f\bar{f}, gg$, at different neutralino masses: $m_{\chi}=$ 6, 10,
50, 100, 500 and 1000 GeV and for pion energies ranging from
$m_{\pi^0}$ to $m_{\chi}$ in 100 equal bins in logarithmic scale. In
order to optimize our numerical calculations, we then obtain the pion
spectrum at neutralino masses and pion energies different from the
ones sampled through a two dimensional numerical interpolation. We
have explicitly checked with the PYTHIA Monte Carlo results that our
interpolation is accurate at the percent level both on the
reconstructed pion yield and on the final gamma--ray spectrum from
$\pi^0$ decay, as given by Eq. (\ref{eq:pi0}).

The contribution to the $\gamma$--ray spectrum from production and
decay of mesons other than pions (mostly $\eta$, $\eta^{\prime}$,
charmed and bottom mesons) and of baryons is usually subdominant as
compared to $\pi^0$ decay. These additional contributions can be
safely neglected (they typically contribute only 
up to 10\% of the total
flux for $E_{\gamma}\lsim$ 1 GeV). A notable exception is given by the
hadronization of $b\bar{b}$ pairs at low production energies, i.e. for
neutralino masses between the production threshold for a $b$--meson
and about 10 GeV. In this case jet flavor conservation leads to the
production of a bottom meson $B=B^0, B^\pm, B^0_s$ with 100 \%
probability. In the PYTHIA code, 75\% of the times the $B$ meson is in
the excited state and decays through $B^* \rightarrow B + \gamma$ with
$m_B^*-m_B \simeq$ 46 MeV. Since the $B^*$ mesons are produced almost
at rest ($m_B \simeq 5.3$ GeV) for $m_\chi\lsim 10$ GeV, they generate
a (slightly boosted) gamma--ray line that dominates the other
contributions below $E_{\gamma} \simeq$ 100 MeV. We have then included
this peculiar contribution to our interpolating procedure \cite{note}.

Neutralino annihilation into lepton pairs can also produce gamma--rays
from electromagnetic showering of the final state leptons. This
process can be dominant for $E_{\gamma}\lsim$ 100 MeV, when the
neutralino annihilation process has a sizable branching ratio into
lepton pairs. In the case of production of $\tau$ leptons, their
semihadronic decays also produce neutral pions, which then further
contribute to the gamma--ray flux. Also these additional contributions
are included in our numerical evaluations, again by modelling the
gamma--ray production with the PYTHIA Monte Carlo for the same set of
neutralino masses quoted above, and by numerically interpolating for
other values of $m_\chi$.

When the neutralino masses are sufficiently large, the annihilation
channels into Higgs bosons, gauge bosons and $t\bar t$ pairs become
kinematically accessible. We compute analytically the full decay chain
down to the production of a quark, gluon or a lepton. The ensuing
$\gamma$--ray spectrum is then obtained by using the results discussed
above for quarks, gluons and leptons, by properly boosting the
differential energy distribution to the rest frame of the annihilating
neutralinos (see e.g. Appendix I in Ref. \cite{pbar_susy} for
details).

\subsection{The geometrical factor}
 
\begin{table*}
\begin{tabular}{|c|c|c|c|c|} \hline
Isothermal & Isothermal & NFW & Moore et al. & $r$--dependent
log--slope Eq.(\ref{eq:alpha}) \\ \hline ~$a=3.5$ kpc~ & ~$a=2.5$ kpc~
& ~$a=25$ kpc~ & ~$a=30$ kpc~ & ~$\alpha=0.142$~ \\ & & ~$r_c=0.01$
pc~ & ~$r_c=0.01$ pc ~& ~$r_{-2}=26.4$ kpc~ \\ & & & & ~$\rho_{-2}=0.035$
GeV cm$^{-3}$~ \\ \hline 18.5 & 42.5 & 184.2 & 10866 & 600 \\ \hline
\end{tabular}
\caption{Values for $I_{\Delta\psi}$ in Eq.(\ref{eq:geometry}) when
different dark matter distributions are assumed (in units of GeV$^2$
cm$^{-6}$ kpc). The angular region of integration $\Delta\psi$ is
defined by the intervals: $|\Delta l| \leq 5^\circ$, $|\Delta b| \leq
2^\circ$. The first two columns refer to an isothermal distribution
with a core $a=3.5$ and 2.5 kpc, respectively. The third and fourth
columns refer to singular DM distributions: a NFW with a scale length
$a=25$ kpc and a Moore et al. profile with a scale length $a=30 $ kpc;
in both cases, the DM profile has a cut--off radius $r_c=0.01$ pc.
The last column refers to the density profile of Eq.(\ref{eq:alpha})
with the parameters of the distribution G1 in Table III of
Ref. \cite{navarro}: $\alpha=0.142$, $r_{-2}=26.4$ kpc and
$\rho_{-2}=0.035$ GeV cm$^{-3}$.  For all these profiles: $\rho_l=0.3$
GeV cm$^{-3}$.}
\label{table:geometry}
\end{table*}

The integral along the line of sight $I(\psi)$ in Eq. (\ref{eq:los})
is the quantity that takes into account the shape of the dark matter
profile. For small values of $\psi$, $I(\psi)$ is very sensitive to
possible enhancement of the density at the GC and therefore large
differences for the different density profiles are expected.

When comparing to experimental data, Eq.(\ref{eq:los}) is averaged
over the telescope aperture--angle $\Delta \psi$:
\beq I_{\Delta\psi} =
\frac{1}{\Delta\psi}\int_{\Delta\psi} I(\psi)~d\psi \, .
\label{eq:geometry}
\eeq
The gamma--ray flux is therefore proportional to $I_{\Delta\psi}$.

In our analysis on the galactic center emission, we will employ data
from the EGRET experiment \cite{egret_gc,egret_gc_source}, whose
angular resolution is given by the longitude--latitude aperture:
$|\Delta l| \leq 5^\circ$, $|\Delta b| \leq 2^\circ$. Table
\ref{table:geometry} shows the values of $I_{\Delta\psi}$ for the
density profiles discussed in Sec. \ref{sec:dmdf}. The effect of
changing the core radius of an isothermal sphere is not negligible: an
increase of a factor of 2.3 is obtained by reducing $a$ from 3.5 kpc
to 2.5 kpc. In the case of singular distributions a small--distance
cut--off of $r_c=0.01$ pc is assumed (inside $r_c$ the density is
assumed to be constant). A NFW profile then gives a flux which is
about 10 times larger than an isothermal sphere with $a=3.5$ kpc. The
very steep Moore et al. profile would produce a flux about 60 times
larger than a NFW. The Navarro et al. \cite{navarro} profile with
$r$--dependent log--slope sits between the NFW and Moore et al. cases:
though not singular, it nevertheless provides a flux which is about 3
times larger than a singular NFW halo. The variability of
$I_{\Delta\psi}$ on the dark matter profile can therefore be as large
as a factor of 600, comparing the very steep Moore et al. profile with
an isothermal sphere with a large core radius. However, the recent
critical analysis on numerical simulation of Ref. \cite{navarro}
implies that a factor of 30 is likely to be a more plausible interval.

In the case of high galactic latitudes, the dependence of
$I_{\Delta\psi}$ on the density profile is much milder.  At these high
latitudes, EGRET identifies a residual gamma--ray flux which is
ascribed to a possible extragalactic component, but could as well be
due to dark matter annihilation in the galactic halo.  We will
consider in our analysis two different angular regions: $|b| >
10^\circ$ with the exclusion of $|l| \leq 40^\circ$ and $10^\circ\leq
|b| \leq 30^\circ$ around the galactic center \cite{egret_extragal}
(region A) and $|b| \geq 86^\circ$ (region B).  Region B has been
considered in the reanalysis of EGRET data by the authors of Ref.
\cite{waxman}, where more stringent limits on the the gamma-ray
residual intensity from the galactic poles have been derived.  The
value of $I_{\Delta\psi}$ for region A is 1.66 for a NFW profile, and
ranges from 1.61 for the isothermal sphere with $a=3.5$ kpc to 1.80
for the $r$--dependent log--slope profile: when pointing away from the
critical behavior of the density profiles at galactic center, the line
of sight integral is almost universal. In the case of region B,
$I_{\Delta\psi}=0.67$ for the NFW distribution, and ranges from 0.62
for the isothermal sphere with $a=3.5$ kpc to 0.69 for the
$r$--dependent log--slope profile.  In both cases, the Moore et
al. profile gives a line--of--sight integral slightly smaller than in
the case of the $r$--dependent log--slope profile.

\subsection{Signal from the galactic center}

\begin{figure*}[t] \centering
\vspace{-20pt}
\includegraphics[width=1.0\columnwidth]{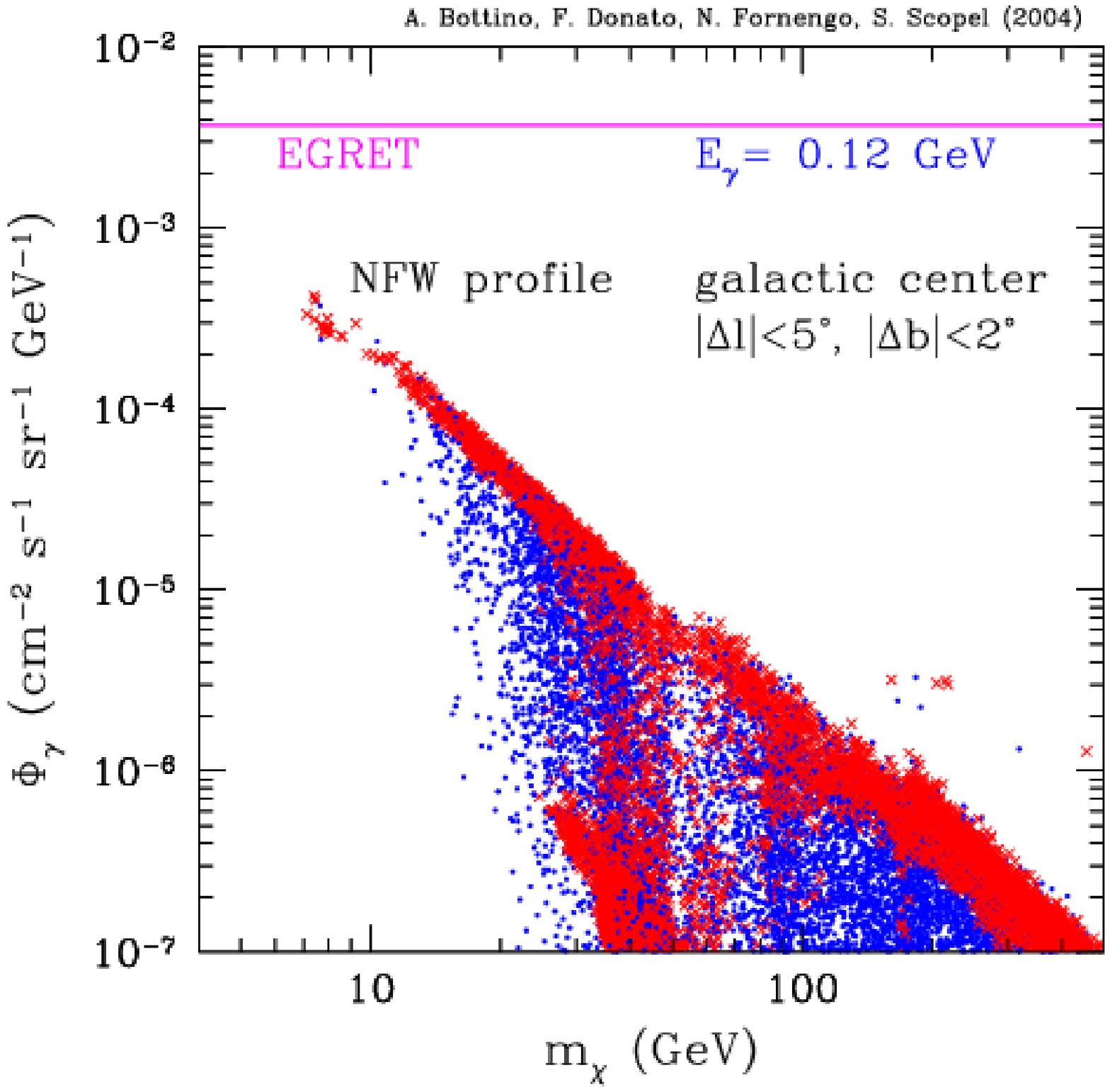}
\includegraphics[width=1.0\columnwidth]{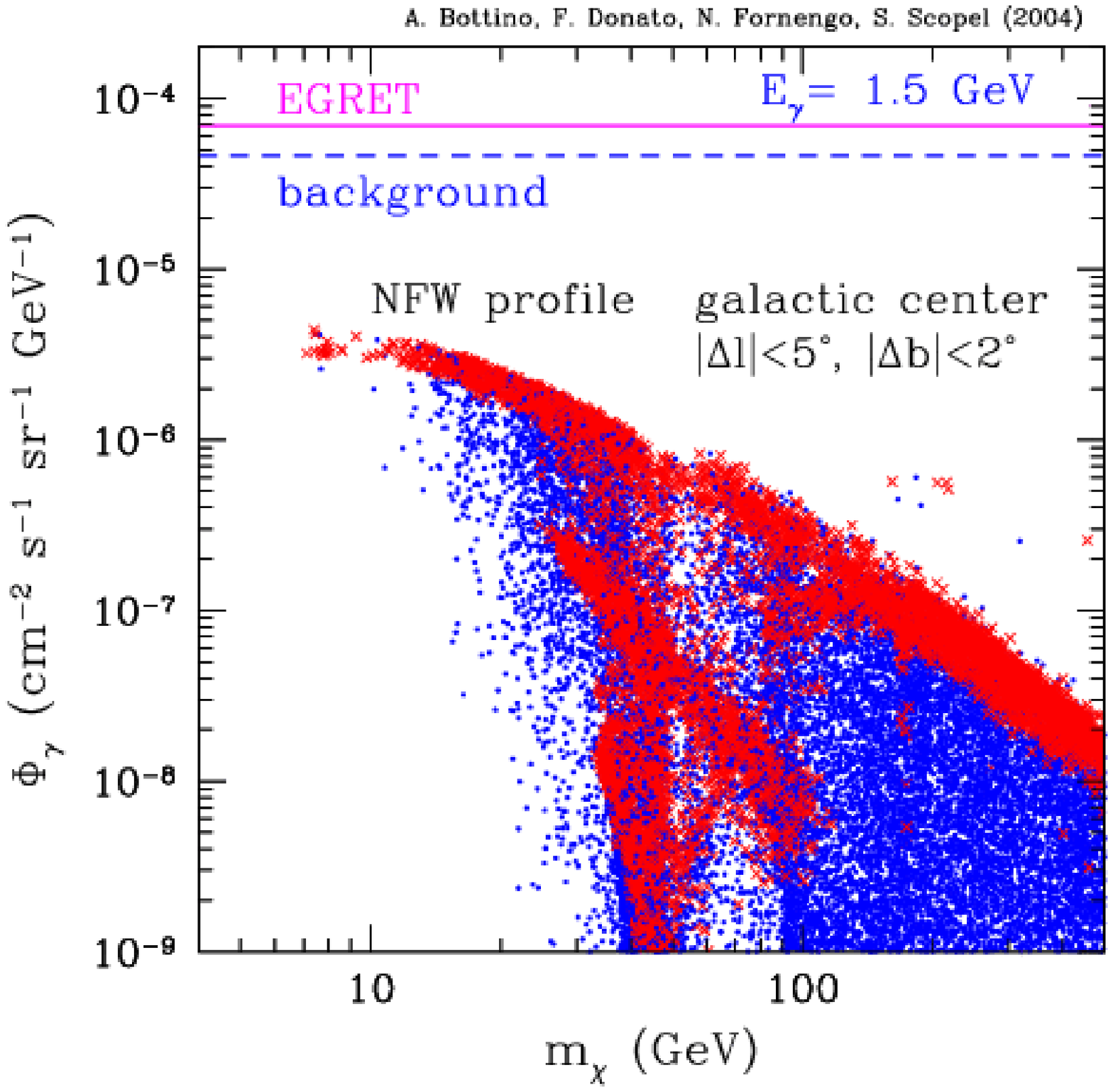}
\includegraphics[width=1.0\columnwidth]{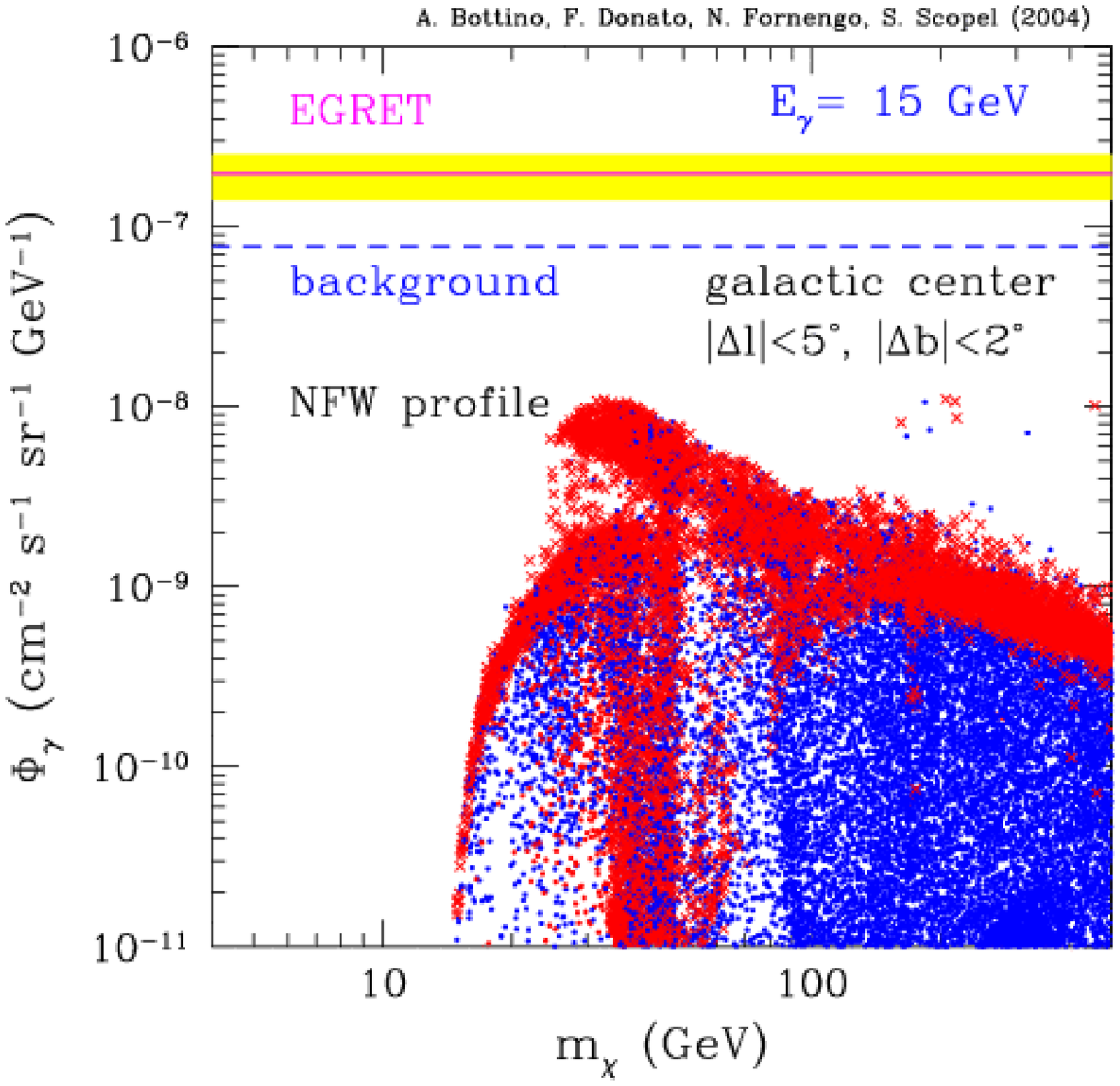}
\vspace{-20pt}
\caption{\label{fig:gammaGC} Gamma--ray flux from the galactic center
inside the angular region $|\Delta l| \leq 5^\circ$, $|\Delta b| \leq
2^\circ$ for a NFW matter density profile. The scatter plots are
derived by a full scan of the parameter space of the supersymmetric
model described in Sec. \ref{sec:susy}.  Crosses (red) and dots
(blue) denote neutralino configurations with $0.095 \leq \Omega_{\chi}
h^2 \leq 0.131 $ and $\Omega_{\chi} h^2 < 0.095$, respectively.  {\sc
Top left:} flux calculated at $E_{\gamma} = 0.12$ GeV; the horizontal
line shows the gamma--ray flux measured by EGRET \cite{egret_gc},
assumed to be compatible with the estimate of the background
\cite{egret_gc}. {\sc Top right:} flux calculated at $E_{\gamma} = 1.5
$ GeV; the solid horizontal line shows the gamma--ray flux measured by
EGRET \cite{egret_gc}, the dashed line is an estimate of the
gamma--ray background \cite{egret_gc}. {\sc Bottom:} flux calculated
at $E_{\gamma} = 15$ GeV; the solid horizontal line shows the
gamma--ray flux measured by EGRET \cite{egret_gc}, the shaded
horizontal band denotes the 1$\sigma$ error bar on the EGRET data and
the dashed line is an estimate of the gamma--ray background
\cite{egret_gc}.}
\end{figure*}

\begin{figure*}[t] \centering
\vspace{-20pt}
\includegraphics[width=1.0\columnwidth]{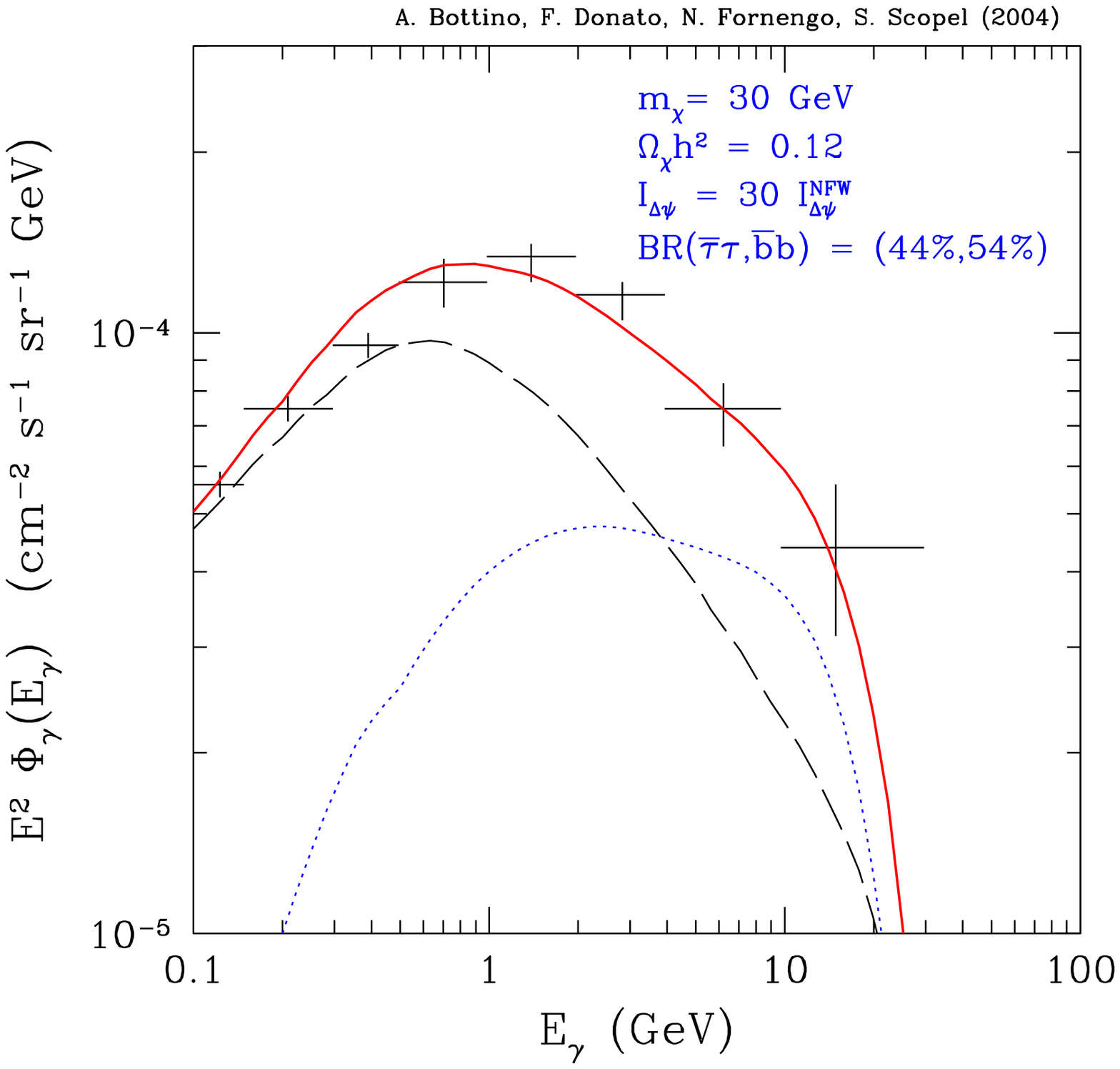}
\includegraphics[width=1.0\columnwidth]{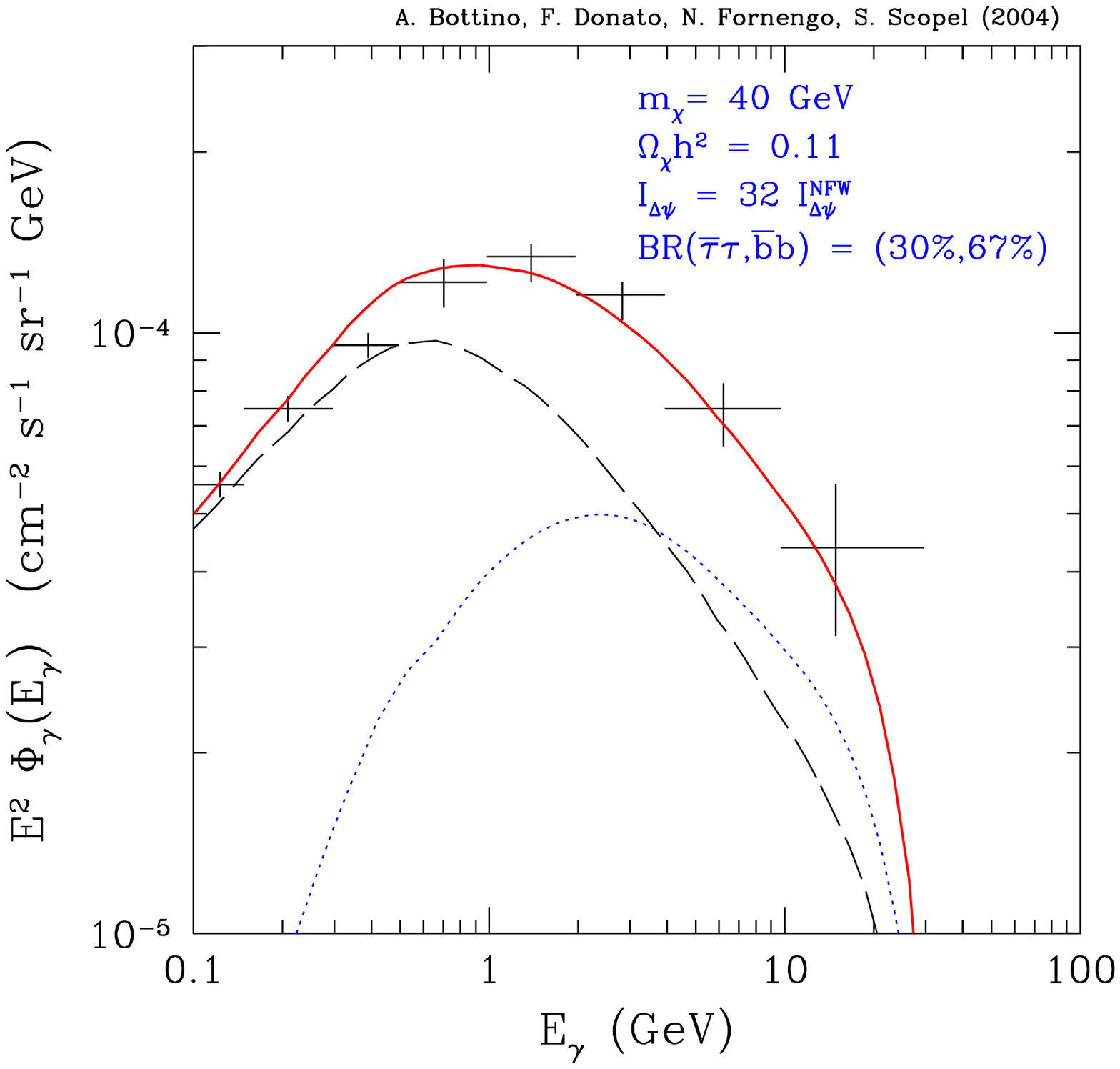}
\vspace{-20pt}
\caption{\label{fig:gammaGC_spectra} Gamma--ray spectra
$\Phi_\gamma(E_\gamma)$, multiplied by $E_\gamma^2$, from the galactic
center inside the angular region $|\Delta l| \leq 5^\circ$, $|\Delta
b| \leq 2^\circ$, as functions of the photon energy. {\sc Left panel:}
the dotted line is the spectrum for a neutralino with mass $m_\chi =
30$ GeV, calculated for a density profile with a factor 30 of
enhancement with respect to the NFW case; the dashed line is the gamma
ray background calculated in Ref. \cite{egret_gc}, reduced by 10\%;
the solid line is the total flux, sum of the supersymmetric signal and
the background; the experimental points are the EGRET data
\cite{egret_gc}. {\sc Right panel:} the same, for $m_\chi = 40$ GeV
and for a density profile with a factor 32 of enhancement with respect
to the NFW case. Both supersymmetric configurations have been selected
from the points shown in Fig. \ref{fig:gammaGC}.  The numbers quoted
in the legend inside parentheses denote the values of the neutralino
annihilation branching ratios into $\bar{b} b$ and $\bar{\tau}\tau$.}
\end{figure*}

Data from low galactic latitudes ($|b|<10^\circ$), including the
galactic center region, have been collected by the EGRET telescope
\cite{egret_gc}. The diffuse gamma--ray flux of the inner galaxy
measured by EGRET shows a possible excess over the estimated
background at energies larger than about 1 GeV.

Clearly a firm assessment of an excess requires a good knowledge of
the standard production of gamma rays in our Galaxy. At the energies
of interest for our analysis - namely, from about 100 MeV to tens of
GeV - the main production mechanism of $\gamma$-rays is the
interaction of cosmic rays (mainly protons and helium nuclei) with the
interstellar medium (atomic and molecular hydrogen, and helium). In
these strong reactions $\pi^0$'s are produced, and hence $\gamma$-rays
via pion decay: $\pi^0 \rightarrow 2\gamma$. The ensuing spectrum has
a bump around 70 MeV, and drops at high energies with an energy power
law which follows the progenitor cosmic ray spectrum ($E^{-\alpha}$,
with $\alpha \sim$ 2.7). Another source of $\gamma$-rays comes from
inverse-Compton scattering of cosmic ray electrons off the
interstellar photons. In particular, energetic electrons may scatter
off the cosmic microwave background, and off the infrared, optical and
ultraviolet radiation arising from stellar activity and dust. The
third radiation component originates from electron bremsstrahlung over
the interstellar medium, which may be partially or even totally
ionized. Bremsstrahlung $\gamma$-rays are mostly important in the low
energy tail. For a full calculation of these three main radiation
components one needs a good knowledge of the physics of cosmic rays
and of the interstellar medium in the region of interest. This is
particularly unlikely when dealing with the galactic center area.

In the literature, several different results have been achieved on the
subject. First of all, the EGRET Collaboration developed a detailed
calculation of the $\gamma$--ray background at the energies of
interest for the detector \cite{egret_gc,egret_calc}: this calculation
shows a clear deficit of $\gamma$-rays towards the GC with respect to
measurements. The excess in the data is apparent for $E_\gamma \gsim
1$ GeV, where the shapes of the spectra of the estimated background and
the data differ significantly. At lower energies the spectral
agreement is instead rather good. Similar conclusions have been drawn
in Refs. \cite{mori} with some different procedure in the calculation
of the background. In this paper a harder, probably unrealistic,
nucleonic spectrum is shown not to be anyway sufficient to explain the
GeV excess. Some modifications towards harder electron and nucleon
spectra are studied in Ref. \cite{strong}, but a satisfactory
agreement with data is not achieved (notice that it is anyway hard to
reconcile these hypothesis with galactic cosmic rays
measurements). The results of all these analyzes favor the
interpretation of the EGRET data in terms of an excess over the
background, mostly at energies above 1 GeV.

However, different assumptions on acceleration \cite{busching} and
diffusion \cite{erlykin} of cosmic ray nucleons, and on the spectral
shape of primary nucleons in the interstellar space \cite{aharonian},
have been proposed and lead to a quite good agreement of the EGRET
measured flux: almost all the spectral features are reproduced by
these calculations. In this case, the EGRET data would be explained in
terms of the standard galactic $\gamma$--ray production.

At present, it is very difficult to favor one model against the
others, both on theoretical and observational basis. This implies that
the uncertainty on the calculation of the galactic $\gamma$-ray flux,
and in particular at the galactic center, is very difficult to
quantify.

Due to these open problems on the determination of the background
component, we will develop our analysis along two paths. First of all
we will discuss whether, and under which conditions, it is possible to set
constraints on low--mass neutralinos from the gamma-ray studies. Then
we will comment on the possibility for low--mass neutralinos to
explain the putative EGRET excess.

The gamma--ray flux from the galactic center inside the angular region
$|\Delta l| \leq 5^\circ$, $|\Delta b| \leq 2^\circ$ for a NFW matter
density profile is shown in Fig. \ref{fig:gammaGC} at three
representative photon energies: $E_\gamma=$ 0.12, 1.5 and 15
GeV. These energies correspond to three energy bins of the EGRET
detector. The scatter plots of the two top panels display a peculiar
funnel shape at small masses. This is due to the fact that the
neutralino flux is bounded from below by the cosmological limit
$\Omega_{\chi} h^2 \leq (\Omega_{CDM} h^2)_{\rm max}$.  This feature
is similar to the one we found in Refs.  \cite{lowneu,lowmass,lowdir},
in connection with neutralino direct detection rates.  The variation
in shape of the scatter plots, when $E_\gamma$ is increased, is easily
understood in terms of Eq.(\ref{eq:flux_gamma}). At $E_\gamma=0.12$
GeV the $m_\chi^{-2}$ behavior is clearly visible. Energies of the
order of 100 MeV are very crucial in offering the possibility to set
limits on the very light neutralino sector. By increasing the photon
energy, the lightest neutralinos do not have enough phase space to
produce photons at this energy (since they annihilate almost at rest
in the Galaxy): therefore the gamma--ray flux at very low masses
becomes progressively depressed, as $E_\gamma$ increases.  At
$E_\gamma=1.5$ GeV a neutralino with a mass of 6 GeV can produce
approximately the same flux as a 10-15 GeV neutralino. At
$E_\gamma=15$ GeV all the neutralinos lighter than 15 GeV obviously do
not produce any photon: at this energy the maximal fluxes are obtained
for neutralinos with masses around 30 GeV.

The first conclusion which can be drawn from Fig. \ref{fig:gammaGC} is
that the supersymmetric model considered here is 
not constrained, at present, by EGRET data for a NFW density
profile. An increase in the flux by a factor of 3.3, as would be in
the case of the $r$--dependent log--slope profile of
Eq.(\ref{eq:alpha}), is also not enough to set limits.

Larger enhancements in the geometrical factor $I_{\Delta\psi}$ over
the NFW case are necessary to set limits. The comparison of the three
panels of Fig. \ref{fig:gammaGC} shows that the limits come from
different energies $E_\gamma$, depending on the neutralino mass. For
very light neutralinos, i.e. $m_\chi \lsim 10$ GeV, the lowest energy
bin is the relevant one. In this case an enhancement of a factor of 6
would allow to raise the predicted fluxes for $E_\gamma=0.12$ GeV at
the level of the EGRET measurement, and therefore to start setting
limits.  For masses in the range $10~{\rm GeV} \lsim m_\chi \lsim
20~{\rm GeV}$ the $E_\gamma=1.5$ GeV bin sets more stringent limits,
at least on a fraction of the supersymmetric models, but only for a
factor of enhancement of at least 15--20 over the NFW case. These
factors are pretty large, even though not as large as the one which
refers to a Moore et al. profile, which is about 60, as discussed
before.  For masses around 30--40 GeV the best limits come from the
highest energy bin $E_\gamma=15$ GeV, where a factor of 20--25 would
allow the fluxes to reach the EGRET data.  In the case of the standard
MSSM, where the neutralino has masses larger than 50 GeV the lowest
energy bins are always less constraining that the higher energy ones,
as can be seen by comparing the different panels of
Fig. \ref{fig:gammaGC}. Instead, the lower energy bins are crucial for
the study of the low--mass neutralinos.  We finally comment that a
Moore et al. profile would make all the fluxes for $m_\chi \lsim 10$
GeV incompatible with the data, but this profile is less likely, as we
discussed above.

\begin{figure*}[t] \centering
\vspace{-20pt}
\includegraphics[width=1.0\columnwidth]{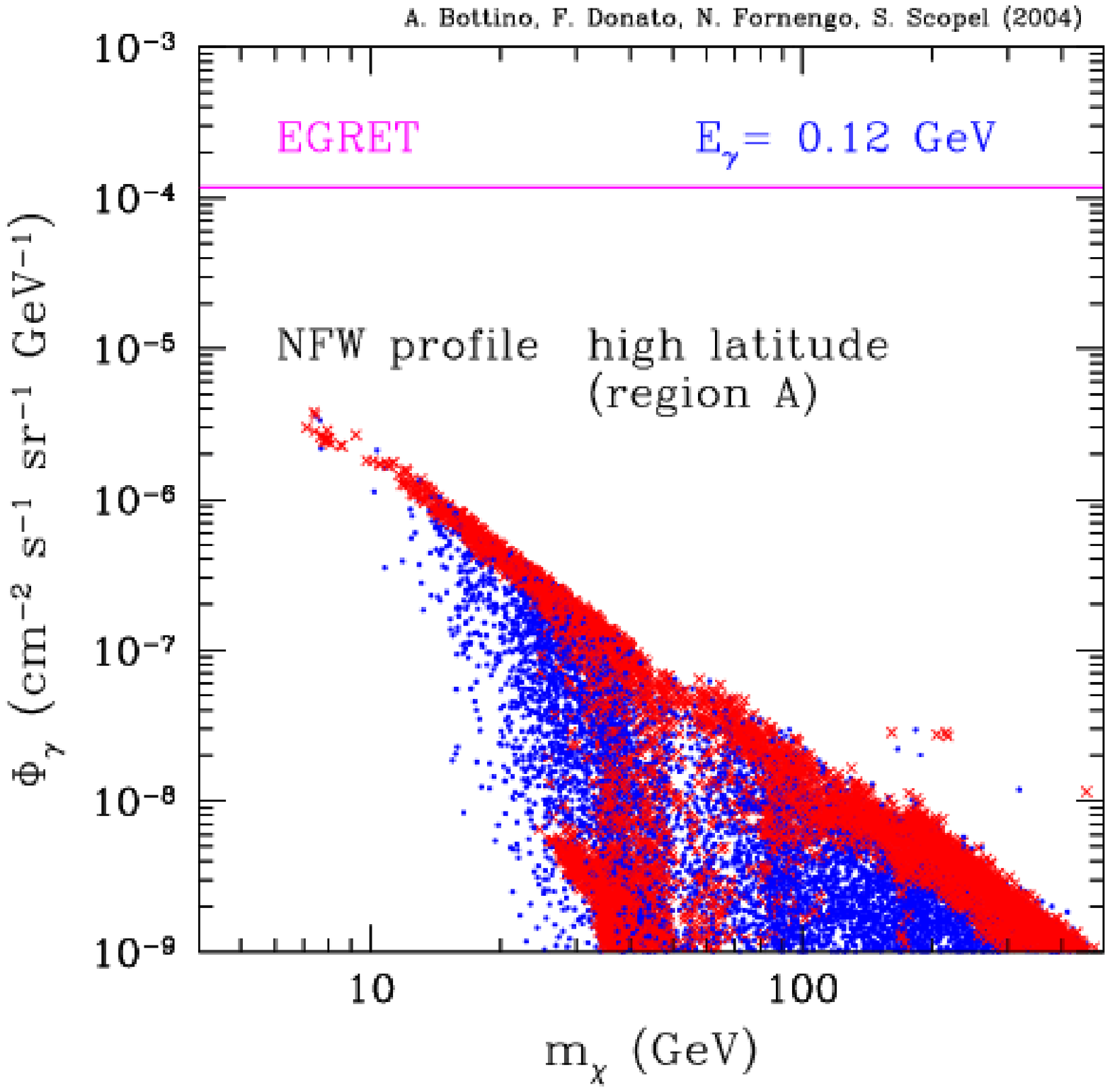}
\includegraphics[width=1.0\columnwidth]{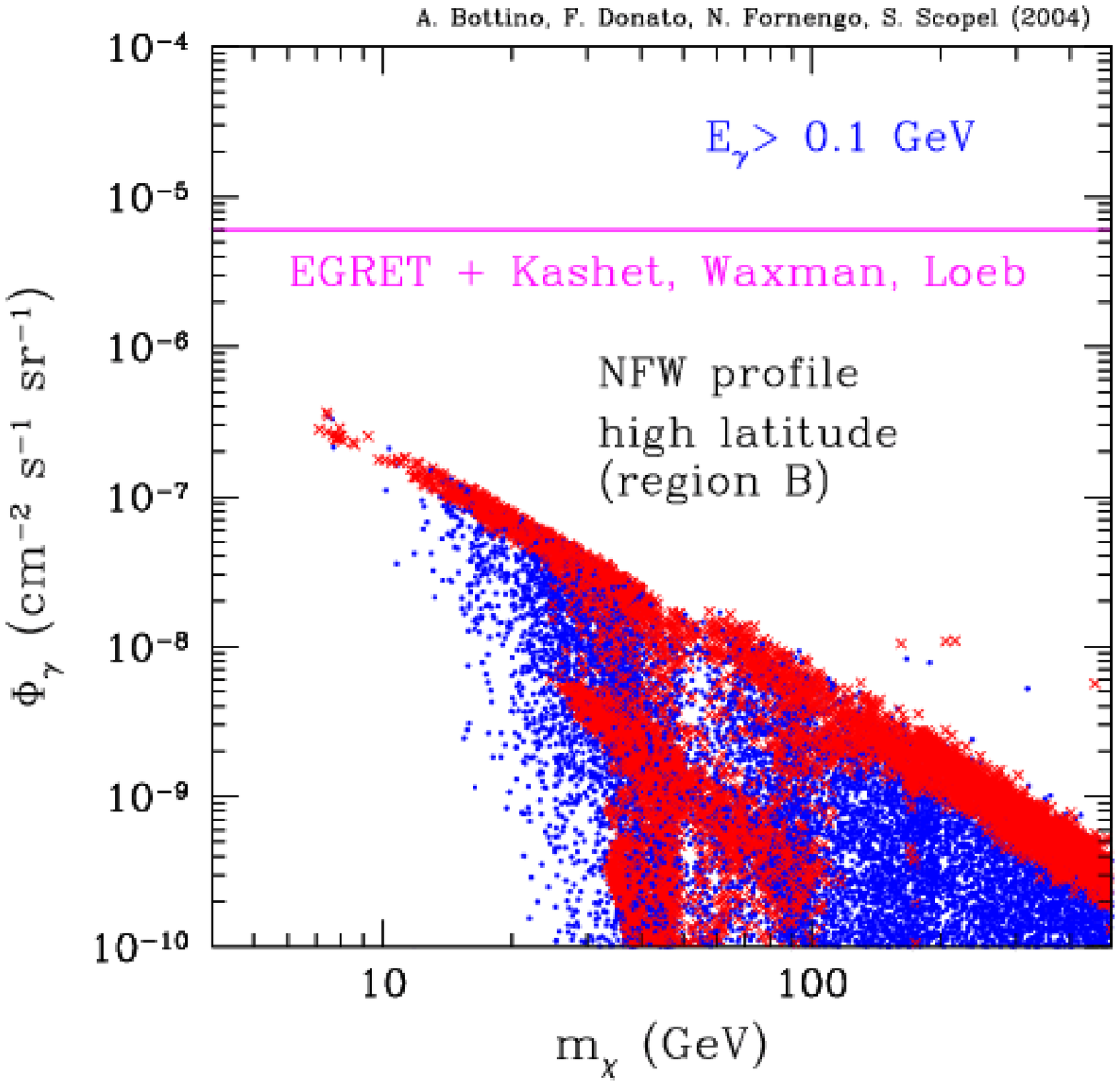}
\vspace{-20pt}
\caption{\label{fig:gammaHL} Gamma--ray flux from galactic
high--latitudes for a NFW matter density profile. The scatter plots
are derived by a full scan of the parameter space of the
supersymmetric model described in Sec. \ref{sec:susy}. Crosses (red)
and dots (blue) denote neutralino configurations with $0.095 \leq
\Omega_{\chi} h^2 \leq 0.131 $ and $\Omega_{\chi} h^2 < 0.095$,
respectively. {\sc Left panel:} flux calculated at $E_{\gamma} = 0.12$
GeV in the high--latitude regions defined by $|b| > 10^\circ$ with the
exclusion of $|l| \leq 40^\circ$ and $10^\circ\leq |b| \leq 30^\circ$
around the galactic center; the horizontal line shows the gamma--ray
residual flux identified by EGRET \cite{egret_extragal}. {\sc Right
panel:} integrated flux for energies above $E_{\gamma} = 0.1$ GeV in
the polar regions defined by $|b| \geq 86^\circ$; the dashed
horizontal line shows the upper limit on a possible residual flux in
the polar regions, obtained in Ref. \cite{waxman}.}
\end{figure*}

Now, let us turn to a brief discussion of the possibility to explain
the EGRET excess by means of low--mass neutralinos. The analysis made
above on the behavior of the gamma--ray fluxes in the three
representative energy bins of Fig. \ref{fig:gammaGC} shows that
neutralinos in the mass range $25~{\rm GeV} \lsim m_\chi \lsim 40~{\rm
GeV}$ are the ones which may have the possibility to fill the excess
in the energy range above 1 GeV, without spoiling the lower energy
behavior of the background which is supposed to have an acceptable
agreement with the data. We show that indeed the low--mass neutralinos
in this mass range are able to explain the EGRET excess in
Fig. \ref{fig:gammaGC_spectra}. In this figure we plot the predicted
gamma--ray spectra for two representative supersymmetric configuration
when, for definiteness, the gamma--ray background as calculated in
Ref. \cite{egret_calc} is assumed. In the left panel of
Fig. \ref{fig:gammaGC_spectra} we show a supersymmetric configuration
with $m_\chi=30$ GeV and a relic abundance in the proper range to
explain dark matter: $\Omega_\chi h^2=0.12$. The dominant annihilation
channels of these low--mass neutralinos are $\chi\chi\rightarrow
\bar{\tau}\tau$ and $\chi\chi\rightarrow \bar{b}b$
\cite{lowneu,lowmass}. Gamma rays coming from annihilation into
$\tau$'s give a harder spectrum as compared to the $b$ channel. In
this representative point the two channels have (approximately) the
same branching ratio: this gives sizable contribution to the
gamma--ray flux in all the energy range from 1 to 10 GeV, which is
where the excess in the EGRET data is more pronounced. By allowing the
background to fluctuate down by 10\% and by using a geometrical factor
$I_{\Delta \psi}$ 30 times larger than in the NFW case, we show that a
pretty good agreement between the total flux and the data can be
obtained. We are not quoting a statistical significance for this
agreement since we are not performing a systematic statistical
analysis here: however, we are interested in showing that, in addition
to the standard MSSM neutralinos with masses larger than 50 GeV
\cite{ullio_morselli}, also low--mass neutralinos have the capability
of explaining the putative EGRET excess. In both cases, low--mass and
standard neutralinos, the values of the line--of--sight integrals
$I_{\Delta \psi}$ which are able to explain the EGRET excess are much
larger than what is provided by a NFW density profile.  However, for
neutralinos in the 30--40 GeV mass range these enhancement factors are
smaller than in the case of heavier neutralinos, due to the
$m_\chi^{-2}$ behavior.

The right panel of Fig. \ref{fig:gammaGC_spectra} shows a second
example, with a neutralino of $m_\chi = 40$ GeV and $\Omega_\chi
h^2=0.11$. The background of Ref. \cite{egret_calc} is again scaled
down by 10\%. In this case the geometrical factor is 32 times the NFW
one. The branching fraction of annihilation into $b$ quarks is larger
than in the previous case.  This enhances the contribution to the
gamma--ray flux in the 1--3 GeV range without spoiling the agreement
at larger energies.

A detailed analysis of the spectral features of the gamma--ray fluxes
produced by neutralino annihilation and their comparison with the
EGRET data is beyond the scope of the present paper and will be
presented elsewhere.

\subsection{Signal from high galactic latitude}

Data from high galactic latitudes have been collected by the EGRET
telescope \cite{egret_extragal}. An analysis of the measurements taken
over the latitudes $|b|>10^o$, and excluding the region $|l|<40^o$ and
$10^o<|b|<30^o$ around the galactic center, has been performed in
Ref. \cite{egret_extragal}. All the identified sources as well as the
components due to the interactions of cosmic rays with the galactic
disk gas have been subtracted \cite{egret_extragal}. The residual
flux, averaged over the considered portion of the sky, has been shown
to be isotropic and well fitted by the power law $\Phi_{HL}^{\rm
EGRET}(E_\gamma) = k (E_\gamma/E_0)^{-\alpha}$, where $k=(7.32
\pm0.34)\cdot 10^{-6}$ photons cm$^{-2}$ sec$^{-1}$ sr$^{-1}$
GeV$^{-1}$, $\alpha=2.10\pm0.03$ and $E_0=451$ MeV. This spectrum is
often referred to as the extragalactic diffuse emission, since no
known source inside the Galaxy seems to be responsible for it. One
possibility is that it is due to unresolved gamma-ray--emitting
blazars. Relying on the analysis by EGRET, one can use the residual
flux as an upper bound to any flux due to exotic sources, including
annihilation of relic neutralinos. Recently, a re-analysis of EGRET
data has been performed in Ref. \cite{waxman}, taking particular care
of the spatial dependence of the observed photons. Working on the
integrated flux and taking into account contributions from several
galactic tracers, the authors of Ref. \cite{waxman} show that the
high-latitude $\gamma$-ray sky exhibits strong galactic features and
could be well accommodated by simple galactic models.  Conservative
constraints have been set on the flux integrated above 100 MeV and
averaged over different sectors of the sky far from the galactic
plane. In this scenario, the room left to an unexplained diffuse flux
-- often considered as an extragalactic background, but also possibly
due to exotic galactic sources -- is much smaller (by a factor of
three, at least) than the one reported in
Ref. \cite{egret_extragal}. Here we consider the upper limit of
Ref. \cite{waxman} on this possible residual, isotropic flux
$I_\gamma$ in the polar region ($|b|>86^0$): $I_\gamma < 0.6 \cdot
10^{-5}$ $ \gamma \; {\rm sec}^{-1} {\rm cm}^{-2} {\rm sr}^{-1}$, and
compare it to our estimates for the $\gamma$-ray flux due to
neutralino annihilation averaged in the same spatial region.

The results for both estimates are shown in Fig. \ref{fig:gammaHL} for
a NFW profile. As in the case of the galactic center emission, also
for high latitudes we do not have constraints on the supersymmetric
parameter space. Contrary to the case of the galactic center region,
for high latitudes the geometrical factor $I_{\Delta\psi}$ is
practically independent of the halo profile, as discussed before.  We
therefore conclude that at present the $\gamma$--ray signals from high
galactic latitudes do not provide any constraint on the supersymmetric
parameter space. The situation could change if further studies will
show that a much bigger fraction of the EGRET measured flux at high
latitudes is due to galactic foreground or when next--generation
experiments will provide further information.

\section{Antiprotons in cosmic rays}
\label{sec:pbar}

\begin{figure*}[t] \centering
\vspace{-20pt}
\includegraphics[width=1.0\columnwidth]{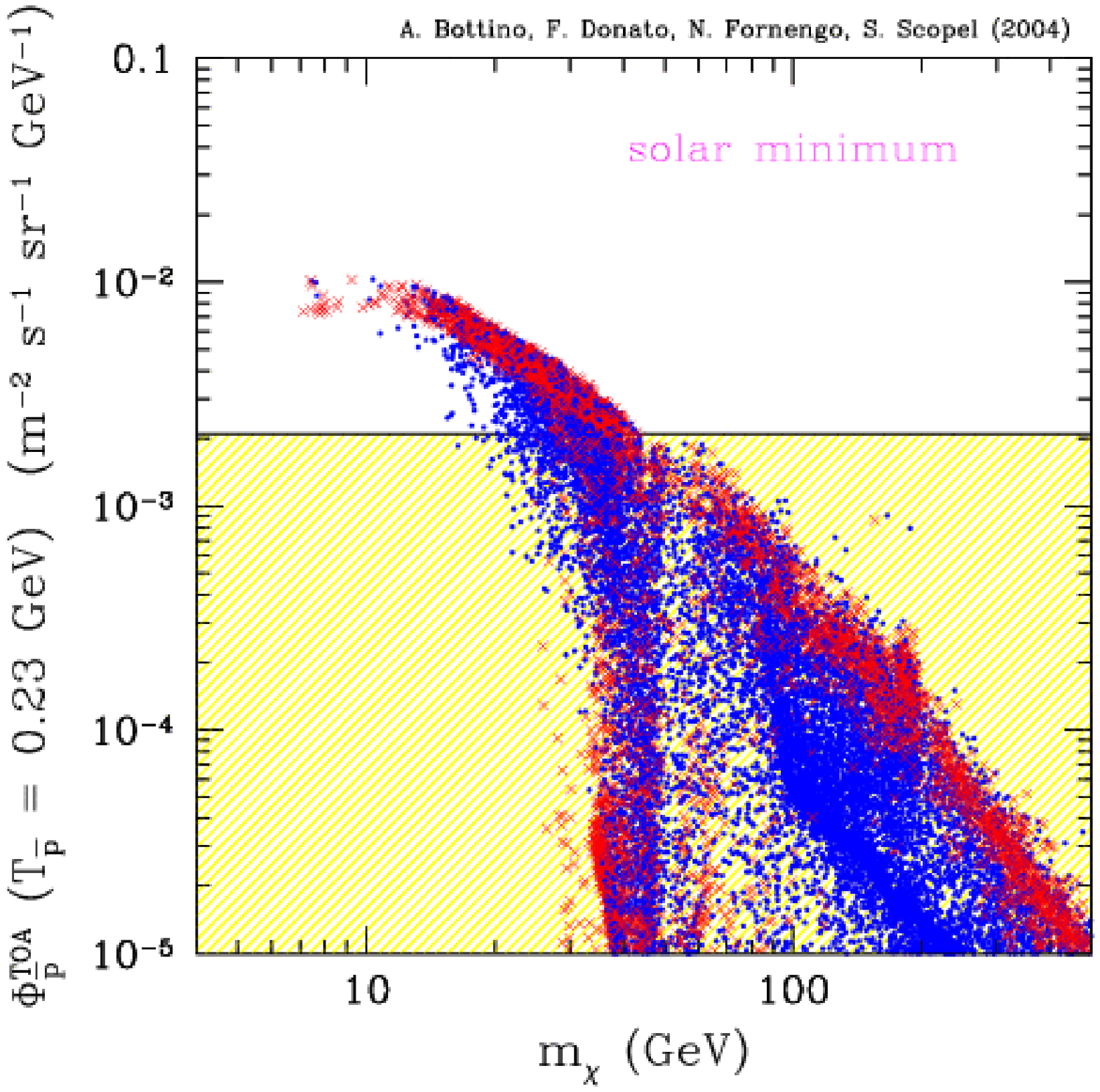}
\includegraphics[width=1.0\columnwidth]{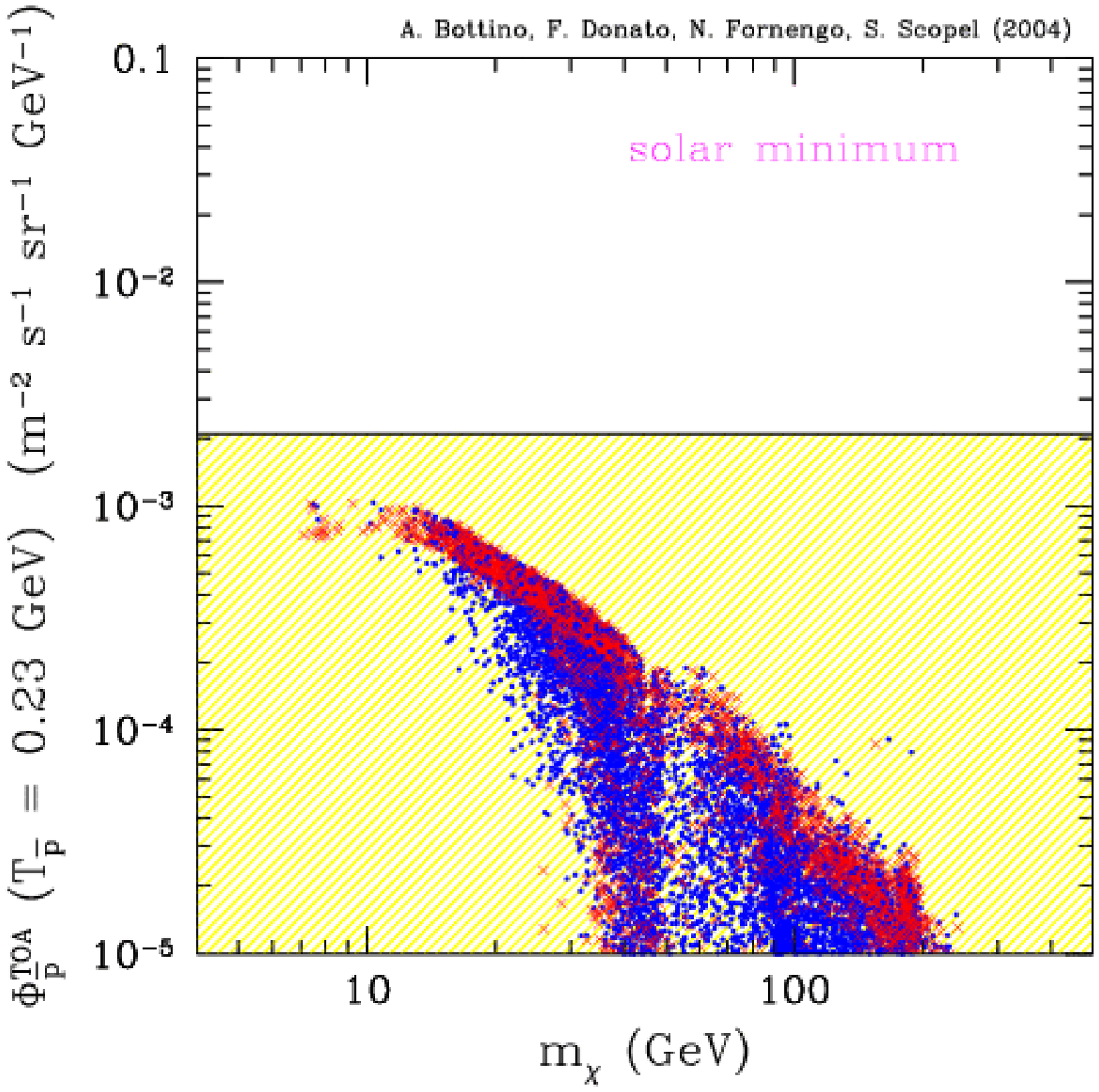}
\vspace{-20pt}
\caption{\label{fig:antiproton} Antiproton flux at $T_{\bar p} = 0.23$
GeV as a function of the neutralino mass, calculated at solar minimum.
The scatter plots are derived by a full scan of the parameter space of
the supersymmetric model described in Sec. \ref{sec:susy}. A spherical
isothermal dark matter density profile has been used. The solar
modulation is calculated at the phase of solar minimum. Crosses (red)
and dots (blue) denote neutralino configurations with $0.095 \leq
\Omega_{\chi} h^2 \leq 0.131 $ and $\Omega_{\chi} h^2 < 0.095$,
respectively. The shaded region denotes the amount of primary
antiprotons which can be accommodated at $T_{\bar p} = 0.23$ GeV
without entering in conflict with the experimental {\sc BESS} data
\cite{bess95-97,bess98} and secondary antiproton calculations
\cite{pbar_sec}. {\sc Left panel:} the best fit set for the
astrophysical parameters is used. {\sc Right panel:} the astrophysical
parameters which provide the most conservative antiproton fluxes are
used.}
\end{figure*}

Like in the case of the gamma--ray flux, the production of antiprotons
from neutralino annihilation results from the hadronization of quarks
and gluons created in the annihilation process
\cite{pbar_history,pbar0,Bottino_Salati,bergstrom,pbar_susy}. The differential
rate per single annihilation, unit volume and time is defined as:
\begin{equation}
q_{\bar p}^{\rm susy}(r,T_{\bar p}) = \, \sigmav \, 
\frac{d N_{\bar p}}{ d T_{\bar p}}
\frac{1}{2}\left( \frac{ \rho_\chi (r)}{m_\chi} \right)^2 ,
\label{eq:source}
\end{equation}
\noindent
where $T_{\bar p}$ denotes the antiproton kinetic energy, $d N_{\bar
p}/ d T_{\bar p}$ (indicated as $g(T_{\bar p})$ in
Refs.\cite{Bottino_Salati,pbar_susy}) is the differential antiproton
spectrum per annihilation event, and the factor 1/2 accounts for the
number of annihilating neutralino pairs.  The spectrum $d N_{\bar p}/
d T_{\bar p}$ is evaluated by means of the Monte Carlo simulations we
already used in Sec. \ref{sec:source_gamma}.

Once antiprotons are produced in the dark halo, they diffuse and
propagate throughout the Galaxy. To describe these processes, we
follow the treatment of Ref. \cite{pbar_susy}, to which we refer for
details. Here we only recall that the propagation of antiprotons has
been considered in a two-zone diffusion model
\cite{PaperI,PaperI_bis,revue}, defined in terms of six free
parameters whose role is to take into account the main physical
processes which affect the propagation of cosmic rays in the Galaxy:
acceleration of primary nuclei, diffusion, convective wind,
reacceleration process and interaction with the interstellar
medium. These free parameters are constrained by comparing the fluxes
of various cosmic ray species calculated in our diffusion model with
observations.  In this regard, the most important observable is the
measured Boron/Carbon ratio ($B/C$), whose analysis within our
diffusion model is presented in Ref. \cite{PaperI_bis}. The parameters
constrained by the $B/C$ measurements have been shown to be compatible
with a series of other observed species
\cite{PaperI_bis,pbar_sec,beta_rad}, further supporting the employed
model. Therefore, in the calculation of the primary antiproton flux we
only use those values for the propagation parameters which provide a
good statistical agreement to the $B/C$ data. One of the main results
obtained in Ref. \cite{pbar_susy} is that the supersymmetric
antiproton flux, when calculated with the selected propagation
parameters, is affected by a large uncertainty. At low energy this
uncertainty reaches almost two orders of magnitude, while it
diminishes to a factor of thirty at higher energies.  Only better and
more complete data on cosmic rays (both stable and radioactive) could
help in reducing this uncertainty.

At variance with the gamma ray signal, the supersymmetric antiproton
flux measured at Earth is almost insensitive to the specific form of
the dark matter distribution function, among those discussed in
Sec. \ref{sec:dmdf}. Indeed, these various distributions differ mainly
at the galactic center, while at solar neighborhood differences are
very mild. Since charged particles, such as antiprotons, suffer
enormous energy redistributions, gains and losses, it is almost
impossible for an antiproton produced around the galactic center to
reach the Earth.  This property was shown in Ref. \cite{origin}, and
quantified in Ref.  \cite{pbar_susy} for the case of a NFW
distribution and an isothermal one.

In the present work, we use directly the results reached in Ref.
\cite{pbar_susy}, where the function
\begin{equation}
C^{\rm prop}_{\rm susy}(T_{\bar{p}})=
\frac{\Phi_{\bar{p}}(r_\odot,T_{\bar{p}})} {\Upsilon
g(T_{\bar{p}})}
\label{eq:C_susy}
\end{equation}
was calculated. In this equation,
$\Phi_{\bar{p}}(r_\odot,T_{\bar{p}})$ is the interstellar antiproton
flux after propagation and $\Upsilon$ is the supersymmetric flux
factor:
\begin{equation}
\Upsilon = \frac{1}{2}  \xi^2 \, \frac{\sigmav}{m_\chi^2}.
\label{eq:upsilon}
\end{equation}
The quantity $C^{\rm prop}_{\rm susy}(T_{\bar{p}})$ may be considered
as a measure of how the source flux $q_{\bar p}^{\rm susy}(r,T_{\bar
p})$ is modified by the propagation of antiprotons in the Galaxy
before reaching the heliosphere.  In the results presented in the
following, we have calculated the antiproton flux
$\Phi_{\bar{p}}(r_\odot,T_{\bar{p}})$ according to
Eq. (\ref{eq:C_susy}), where the $C^{\rm prop}_{\rm
susy}(T_{\bar{p}})$ function has been taken directly from
Ref. \cite{pbar_susy} for a few representative combinations of the
propagation parameters and source spectra $g(T_{\bar p})$.  We have
calculated the quantities entering the factor $\Upsilon$ as described
in Sec. \ref{sec:susy}.

\subsection{Secondary antiprotons}

Antiprotons in the Galaxy are also produced via standard
interactions. Proton and helium cosmic rays interact with the
interstellar hydrogen and helium nuclei, producing quarks and gluons
that subsequently can hadronize into antiprotons. A calculation of
this secondary antiproton flux has been done in Ref.  \cite{pbar_sec},
to which we refer for details. Here we only emphasize the main results
obtained in that work: i) The antiproton flux has been evaluated
consistently by employing the propagation parameters as derived from a
full and systematic analysis on stable nuclei \cite{PaperI}; ii) This
secondary antiproton flux is in very good agreement with the data
taken from the experiments {\sc bess} \cite{bess95-97,bess98}, {\sc
ams} \cite{ams98}, {\sc caprice} \cite{caprice} (see Fig. 14 in
Ref. \cite{pbar_susy}); iii) The uncertainty on the final flux due to
propagation is about 20\%, with a slight dependence on the energy.
Another important source of uncertainty of order 20\%-25\% resides in
the nuclear production cross sections, in particular when considering
the interactions over the interstellar helium.

\subsection{Constraints on a primary antiproton source?}

As discussed above, the secondary antiproton flux already provides a
satisfactory agreement with current experimental data, and then no
much room is left to primary contributions.  This situation suggests
that antiproton data could be used to place significant constraints on
supersymmetric parameters.  However, one has to notice that, as shown
in \cite{pbar_susy}, the supersymmetric primary flux is affected by
uncertainties much larger than those related to the secondary flux.
This is due to the fact that the sources of the latter are located in
the galactic disk. On the contrary, the relic neutralinos are expected
to be distributed in the whole galactic halo and then produce an
antiproton flux much more sensitive to the astrophysical parameters.

To show quantitatively how the experimental data could constrain the
supersymmetric parameters, in Fig. \ref{fig:antiproton} we display the
antiproton flux evaluated at $T_{\bar p}=0.23$ GeV for a full scan of
our supersymmetric model described in Sec. \ref{sec:susy}.  As
expected, the scatter plot is prominent at small masses.  Furthermore,
it is remarkable that for $m_\chi \lsim$ 25 GeV the scatter plot is
funnel-shaped. The reason is the same as the one given above in
connection with Fig. \ref{fig:gammaGC}.  The two panels of
Fig. \ref{fig:antiproton} correspond to two different sets of the
propagation parameters. One, used in the left panel, is the set giving
the best fit to the $B/C$ ratio, while the other, hereby denoted as
the conservative set and used in the right panel provides the lowest
(secondary and primary) antiproton fluxes.  A spherical cored
isothermal distribution for dark matter has been used.  However, as
mentioned before, a different choice does not significantly modifies
the scatter plots.  The shaded region denotes the amount of primary
antiprotons which can be accommodated at $T_{\bar p} = 0.23$ GeV
without entering in conflict with the {\sc BESS} experimental data
\cite{bess95-97,bess98} and secondary antiproton calculations
\cite{pbar_sec}.

From the right panel of Fig. \ref{fig:antiproton} we conclude that,
within the current astrophysical uncertainties, one cannot derive any
constraint on the supersymmetric parameters, if one assumes a very
conservative attitude in the selection of the propagation parameters.
It is worth noticing that even within this choice, some supersymmetric
configurations at very small masses are close to the level of
detectability.  As a further comment on the left panel of
Fig. \ref{fig:antiproton}, we wish to stress that any further
breakthrough in the knowledge of the astrophysical parameters would
allow a significant exploration of small mass configurations, in case
the conservative set of parameters is excluded.  Should the effect of
antiproton propagation turn out to be equivalent to the one obtained
with the best fit set, the analysis of cosmic antiprotons would prove
quite important for exploring very light neutralinos. This is
particularly true for neutralino masses below $\lsim$ 15 GeV, in view
of the typical funnel shape displayed in the scatter plots.

\section{Upgoing muons at neutrino telescopes}
\label{sec:nu}

\begin{figure*}[t] \centering
\vspace{-20pt}
\includegraphics[width=1.0\columnwidth]{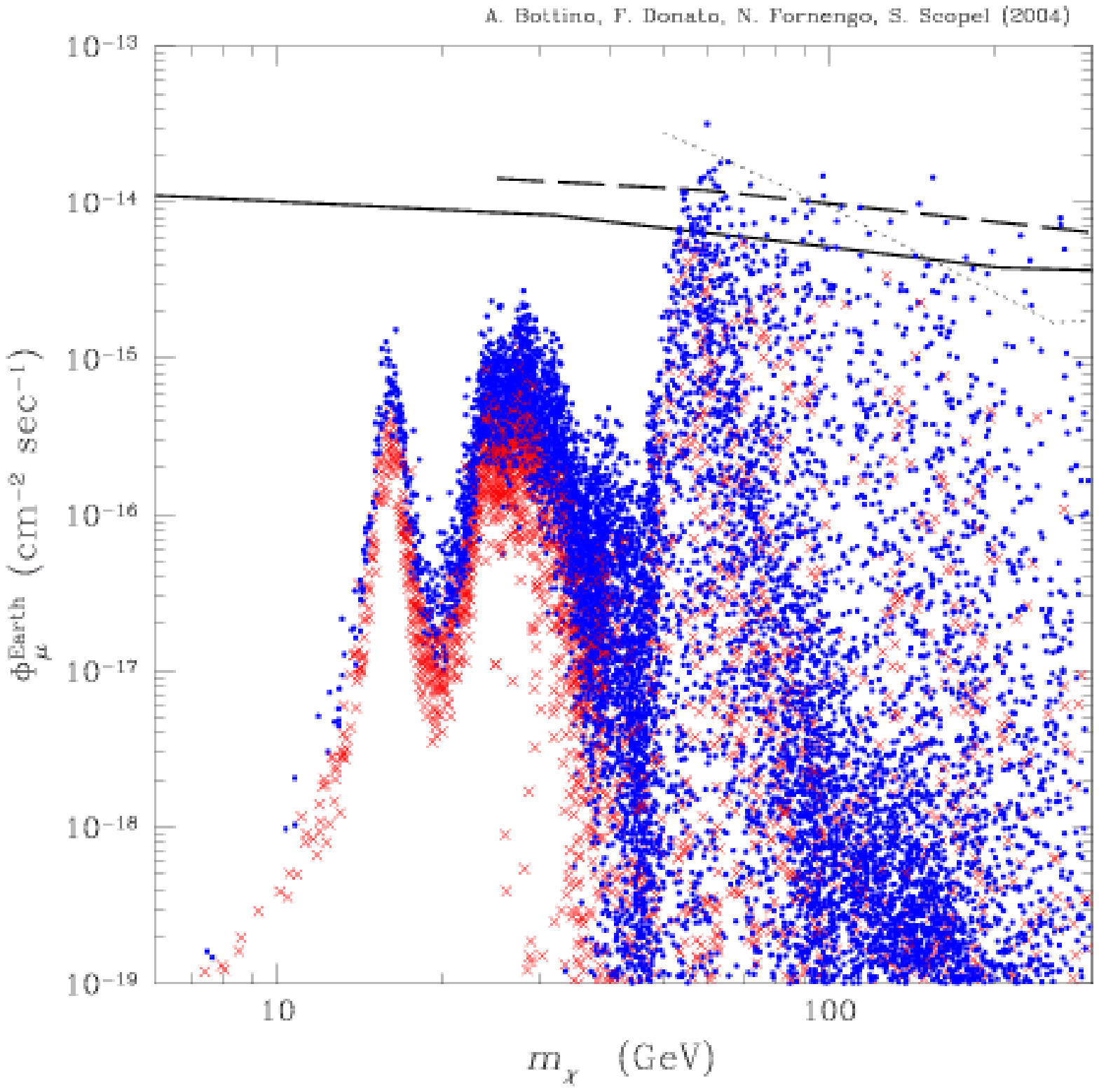}
\includegraphics[width=1.0\columnwidth]{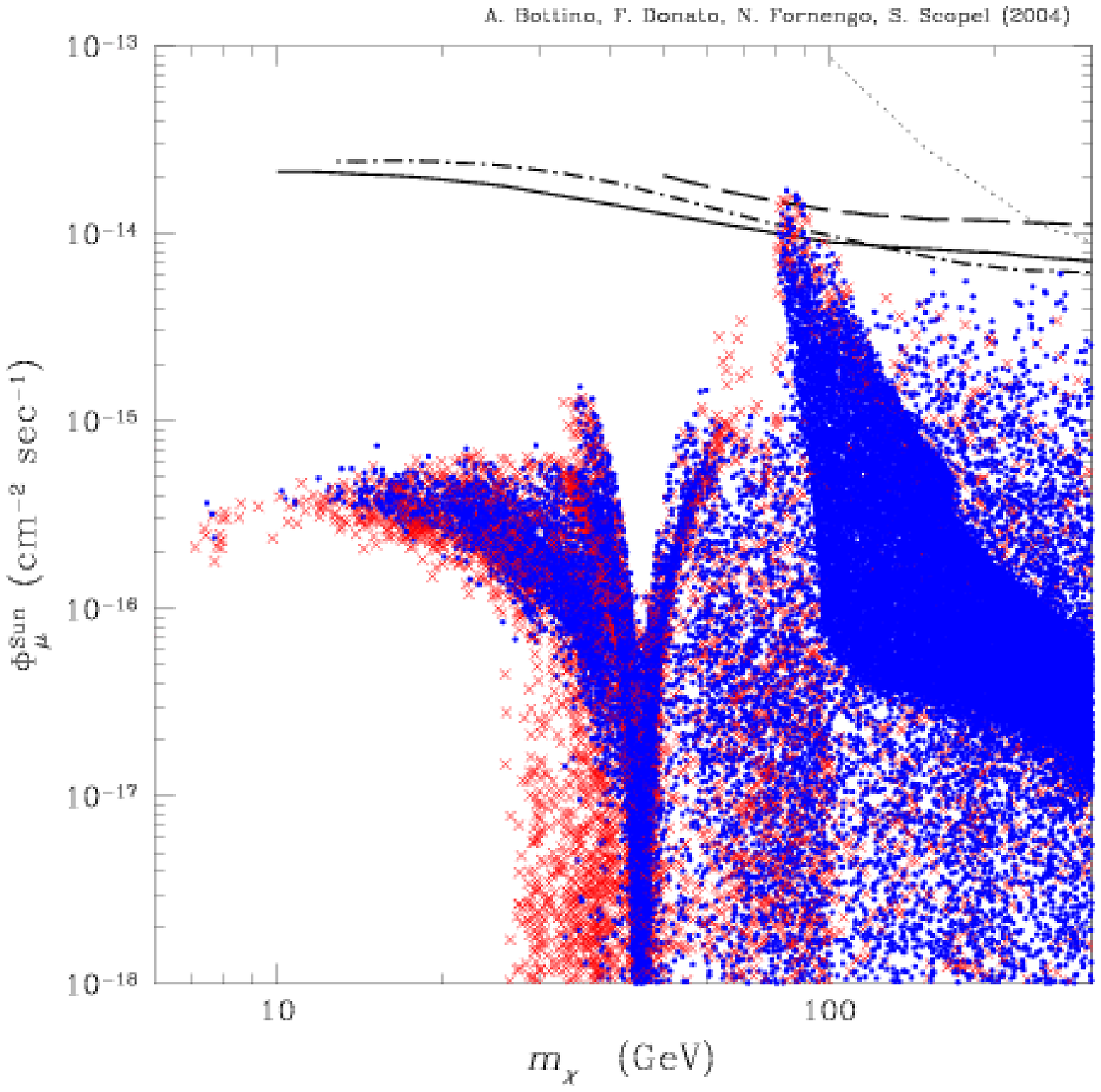}
\vspace{-20pt}
\caption{\label{fig:neutrinos} Flux of upgoing muons as a function of
the neutralino mass. The scatter plots are derived by a full scan of
the parameter space of the supersymmetric model described in
Sec. \ref{sec:susy}.  Crosses (red) and dots (blue) denote neutralino
configurations with $0.095 \leq \Omega_{\chi} h^2 \leq 0.131 $ and
$\Omega_{\chi} h^2 < 0.095$, respectively. {\sc Left panel:} signal
from the Earth; the solid, dashed and dotted lines denote the
experimental upper limits from SuperKamiokande \cite{sk}, MACRO
\cite{macro} and AMANDA \cite{amanda}, respectively. {\sc Right
panel:} signal from the Sun; the solid, dashed, dot-dashed and dotted
lines denote the experimental upper limits from SuperKamiokande
\cite{sk}, MACRO \cite{macro}, Baksan \cite{baksan} and AMANDA
\cite{amanda}, respectively.}
\end{figure*}

Indirect evidence for WIMPs in our halo may be obtained at neutrino
telescopes by measurements of the upgoing muons, which would be
generated by neutrinos produced by pair annihilation of neutralinos
captured and accumulated inside the Earth and the Sun
\cite{others_mu,noi_mu}.  The process goes through the following steps:
capture by the celestial body of the relic neutralinos through a
slow--down process due essentially to neutralino elastic scattering
off the nuclei of the macroscopic body; accumulation of the captured
neutralinos in the central part of the celestial body;
neutralino--neutralino annihilation with emission of neutrinos;
propagation of neutrinos (we have included the $\nu_\mu$--$\nu_\tau$
vacuum oscillation effect with parameters: $\Delta m^2=3\times
10^{-3}$ eV$^2$, $\sin\theta=1$) and conversion of their $\nu_{\mu}$
component into muons in the rock surrounding the detector; propagation
and detection of the ensuing upgoing muons in the detector.

The various quantities relevant for the previous steps are calculated
here according to the method described in
Refs. \cite{noi_mu}, to which we refer for further details.
We just recall that the neutrino flux due to the annihilation
processes taking place in a distant source like the Sun, as a function
of the neutrino energy $E_{\nu}$, is given by
\beq
\frac{dN_\nu}{dE_\nu}=\frac{\Gamma_A}{4\pi d^{\,2}}\sum_{F,f}
B^{(F)}_{\chi f}\frac{dN_{f \nu}}{dE_\nu},
\label{eq:nflux}
\eeq
where $\Gamma_A$ is the annihilation rate inside the macroscopic body
\cite{griest}, $d$ is the distance from the source, $F$ denotes the
$\chi$--$\chi$ annihilation final states, $B^{(F)}_{\chi f}$ denotes
the branching ratios into heavy quarks, $\tau$ leptons and gluons in
the channel $F$; $dN_{f \nu}/dE_{\nu}$ is the differential
distribution of the neutrinos generated by the hadronization of quarks
and gluons and the subsequent hadronic semileptonic decays.  The
annihilation rate is given by $\Gamma_A = C/2 ~{\rm tanh}^2
(t_0/\tau_A)$ \cite{griest}, where $t_0$ is the age of the macroscopic
body, $\tau_A = (C C_A)^{-1/2}$, $C_A$ is the annihilation rate
proportional to the neutralino--neutralino annihilation cross--section
and $C$ denotes the capture rate per effective volume of the body.
The capture rate $C$ is calculated here as in
Refs. \cite{gould,helio}, the WIMP velocity distribution in the
galactic inertial frame being described by a Maxwellian function with
a dispersion velocity of 270 km s$^{-1}$. We recall that the capture
rate by the Earth is favored when the WIMP mass is close to the
nuclear mass of one of the main chemical components of the Earth:
Oxygen, Magnesium, Silicon in the mantle and Iron in the core
\cite{gould}.  We have neglected the contributions of the light quarks
directly produced in the annihilation process or in the hadronization
of heavy quarks and gluons, because these light particles stop inside
the medium (Sun or Earth) before their decay.  For the case of the Sun
we have also considered the energy loss of the heavy hadrons in the
solar medium.

\subsection{Neutrinos from the center of the Earth}

In the left panel of Fig. \ref{fig:neutrinos} we show the expected
upgoing muon flux integrated for $E_\mu>$1 GeV, as a function of
$m_\chi$, and compared to the present experimental upper bounds on the
same quantity from the experiments Superkamiokande \cite{sk}, MACRO
\cite{macro}, and AMANDA \cite{amanda}.  For $m_\chi\lsim$ 40 GeV the
signal from the Earth presents several peaks due to resonant capture
on Oxygen, Silicon and Magnesium (we recall that the process of
capture on Earth is driven by the coherent neutralino--nucleus cross
section). These elements are almost as abundant in Earth as Iron,
which is the most important target nucleus for neutralino capture at
higher masses. The dip at $m_\chi\simeq M_Z/2$ is due to the rescaling
prescription of Eq. (\ref{eq:rescal}), since the resonance in the
$Z$--exchange annihilation cross section reduces the relic abundance
$\Omega_\chi h^2$. Moreover, for $m_\chi\lsim$ 25 GeV, the branching
ratio $B^{(\tilde{\tau})}_{\chi \tau}$ to the $\tau \bar{\tau}$ final
state, which is the one with the highest neutrino yield per
annihilation, is suppressed. This last property is due to the fact
that, in this range of $m_\chi$, the final state to $b \bar{b}$ in the
annihilation cross section is required to be the dominant one in order
to keep the relic abundance $\Omega_\chi h^2$ below its cosmological
upper bound \cite{lowmass}. This, together with the fact that lower
$\chi$ masses imply softer $\nu$ spectra that produce less $\mu$'s
above threshold, explains why the up-going muon signal expected for
light neutralinos ($m_\chi\lsim$ 50 GeV) turns out to be below the
level reached at higher masses.

It is worth noting that substantial modifications to the standard
Maxwellian velocity distribution in the solar neighborhood have been
considered in recent years, with two conflicting models. Damour and
Krauss \cite{dk} proposed the existence of a solar-bound population,
generated by WIMPs which scattered off the Sun surface and were then
set, by perturbations from other planets, into orbits crossing the
Earth, but not the Sun. The existence of this low-speed population
would make the capture of WIMPs by the Earth very efficient with an
ensuing dramatic enhancement of the up-going muon flux as compared to
the standard case \cite{bdeku}. On the contrary, Gould and Alam
\cite{ga} used arguments based on calculations of asteroid
trajectories to conclude that solar-bound WIMPs could evolve quite
differently, inducing a significant suppression in the up-going muon
flux from the Earth, as compared to the standard Maxwellian case, for
WIMP masses larger than 150 GeV. A very recent re--analysis of this
problem \cite{le} supports the conclusions of Gould and Alam, though
with a less dramatic suppression effect. Our results have been derived
using the standard Maxwellian distribution. The results of
Ref. \cite{le} would not significantly alter our conclusion on the
detectability of light neutralinos. The maximal fluxes are obtained
for resonant capture on $O$, $Si$ and $Mg$ nuclei in the mantle: in
this situation no suppression occurs. For neutralino masses away from
the resonant condition, a reduction factor up to 0.8 for $m_{O} \lsim
m_\chi \lsim m_{Fe}$ and up to 2--3 for $m_\chi \lsim 10$ GeV is
possible \cite{le}: however, in these cases, the upgoing muon flux is
already very depressed, as it is shown in Fig. \ref{fig:neutrinos},
left panel.  In conclusion, using the standard Maxwellian
distribution, the present measurements of up-going muons from the
Earth put some constraints on neutralino configurations for masses
above 50 GeV.  For lighter neutralinos, explorations by neutrino
telescopes would require a substantial increase in sensitivity while
keeping a low energy threshold (close to 1 GeV). This in turn would
imply a sizable extension of the telescope and a dense array of
photomultipliers, which is certainly feasible, but very expensive.

\subsection{Neutrinos from the Sun}

In the right panel of Fig. \ref{fig:neutrinos} we show the up-going
muon flux expected from the Sun, integrated for $E_\mu>$1 GeV, as a
function of $m_\chi$.  The signal is compared to the present
experimental upper bounds on this flux coming from the experiments
Superkamiokande\cite{sk}, MACRO\cite{macro}, AMANDA\cite{amanda} and
Baksan\cite{baksan}.  Also in this case the signal level turns out to
be suppressed for $m_\chi\lsim$ 50 GeV as compared to what is obtained
at higher masses, the reasons of this behavior being the same as in
the case of the Earth. On the other hand, the enhancement of the
signal at $m_\chi\sim m_W$ is due to a peculiar behavior of the
neutralino--nucleon spin--dependent cross section, which drives the
neutralino capture in the Sun (mainly on hydrogen).  This cross
section reaches its maximum whenever the $Z$--exchange channel
dominates, and this requirement is verified when the neutralino--$Z$
coupling, proportional to the combination $a_3^2-a_4^2$, is maximal
\cite{spin}. By numerical inspection we have verified that this last
quantity is significantly peaked for $m_\chi\sim M_W$. In this range
of masses the annihilation channel to $W^{+} W^{-}$ opens up and
dominates the annihilation cross section.  We conclude here that
investigations of light neutralinos by up-going muons from the Sun do
not provide favourable prospects.

\section{Conclusions and Perspectives}
\label{sec:conclusions}

We have considered the most relevant indirect strategies for detecting
the presence of relic neutralinos in our Galaxy through the products
of their self-annihilation.  This includes annihilations taking place
directly in the galactic halo or inside celestial bodies (Earth and
Sun). Our investigation has been performed in the frame of an
effective supersymmetric model at the electroweak scale with no
assumption on gaugino mass universality at the GUT scale.  The range
of the neutralino mass taken into consideration brackets a wide
interval, from 6 GeV up to 500 GeV. While the low extreme is decided
by the lower bound of 6 GeV, established in
Ref. \cite{lowneu,lowmass}, the upper extreme of 500 GeV is chosen
only for convenience.  Actually, no model-independent upper limit for
the neutralino mass is available, apart for a generic value of order
of 1 TeV, beyond which the raison d'\^etre of supersymmetry fades
away.  Though our calculations span over the wide range of the
neutralino masses recalled above, our discussions were focused on
light neutralinos, {\it i.e.} neutralinos with masses below 50 GeV:
this value corresponds to the lower bound of $m_\chi$ when
gaugino-mass unification is assumed. Indeed, light neutralinos are
rarely considered in the literature, though their properties are quite
interesting, as we already proved in connection with their
cosmological properties and their detectability by current experiments
of direct WIMP search \cite{lowneu,lowmass,lowdir}. Thus the present
paper is the natural continuation of our previous investigations on
neutralinos of small mass.  Different galactic dark matter
distributions have been considered, from the cored isothermal one to
the profiles obtained by numerical cosmological simulations in
Refs. \cite{nfw,moore}, including also the most recent ones of
Ref. \cite{navarro}.

Now we summarize the main results of the present paper:

\begin{itemize}

\item For the $\gamma$ rays we have considered separately fluxes from
the galactic center and high latitude regions, and compared our
predictions with the EGRET data. Our numerical results have been
provided employing as a reference DM distribution the NFW profile.  We
have shown that in this case the EGRET data at all angles do not put
any constraints on the supersymmetric flux. The minimum gap between
the theoretical predictions and the data occurs at light masses and is
of almost one order of magnitude. We have discussed by how much this
gap changes in terms of the dark matter distribution. This variation
is relevant only for signals coming from the galactic center. For this
sector, we have shown that using the $r$--dependent log--slope
distribution of Ref. \cite{navarro} the gap between data and
supersymmetric fluxes is reduced by a factor of 3 with respect to the
NFW profile. Only profiles as steep as the Moore et al. one,
disfavored by recent simulations, could exclude some light neutralino
configurations. We have also shown that neutralinos of masses around
30--40 GeV could explain the EGRET excess in case of a significant
enhancement effect as compared to the NFW distribution. A general
word of caution concerns the fact that the background due to
conventional cosmic rays production mechanisms still suffers from
sizeable uncertainties.

\item We have shown that in the case of cosmic antiprotons, no
constraint on the supersymmetric parameters can be derived, if one
assumes a very conservative attitude in the selection of the
propagation parameters. However, it is remarkable that indeed the
signal at very small masses is close to the level of
detectability. Some breakthrough in the knowledge of the astrophysical
parameters could allow a significant exploration of small mass
configurations.  This is particularly true for neutralino masses below
about 15 GeV, in view of the typical funnel shape displayed in the
scatter plots.
  
\item The present measurements of up-going muons from the center of
the Earth put some constraints on neutralino configurations for masses
above 50 GeV.  For lighter neutralinos, explorations by neutrino
telescopes require a substantial increase in sensitivity with an
energy threshold close to 1 GeV. Investigations of light neutralinos
by up-going muons from the Sun are very disfavored.

\end{itemize}

We wish to recall that, according to the measurements of the HEAT
Collaboration \cite{heat}, the spectrum of the positron component of
cosmic rays shows some enhancement between 7 and 20 GeV. This is only
a mild effect which, as shown in Ref. \cite{heat}, could be explained
in terms of conventional secondary production
mechanisms. Alternatively, some authors have interpreted this effect
as a deviation from a pure secondary flux, which could be generated by
neutralino self-annihilation \cite{kane,be}. This hypothesis, to be a
viable one, requires that the neutralino flux is enhanced by a sizable
factor. Since the measured positrons must be created in a region
around the Earth of a radius of a few kpc \cite{origin}, an
enhancement would imply a significant dark matter overdensity in that
region, with implications for the $\bar p$ signal. This scenario
appears strongly model dependent, and as such not suitable for setting
constraints to supersymmetric parameters.

Antideuterons in space as a signal of neutralino self-annihilation,
which were shown in Ref.\cite{dbar} to be a very promising
investigation means, will be considered in a forthcoming paper.

\acknowledgments 
We acknowledge Research Grants funded jointly by the
Italian Ministero dell'Istruzione, dell'Universit\`a e della Ricerca
(MIUR), by the University of Torino and by the Istituto Nazionale di
Fisica Nucleare (INFN) within the {\sl Astroparticle Physics Project}.
We thank the Referee for suggesting  the inclusion of the upper
bound on the $BR(B_s \rightarrow \mu^+ + \mu^-)$ decay in our analysis.



\begin{thebibliography}{99}
\bibitem{note}  More details about our evaluation of the $\gamma$
spectrum from neutralino self--annihilations will be presented in a
separate paper.
\end{thebibliography}
\end{document}